\documentclass[aps,prd,onecolumn,nofootinbib,11pt,eqsecnum,preprintnumbers,superscriptaddress]{revtex4-2}
\usepackage{graphicx}
\usepackage{amssymb}
\usepackage{mathtools}
\usepackage{enumitem}

\usepackage{amsmath,amsthm}
\theoremstyle{plain}
\newtheorem{thm}{Theorem}[section]

\theoremstyle{definition}
\newtheorem{ass}[thm]{Assumption}

\theoremstyle{remark}

\theoremstyle{plain}
\newtheorem*{thm*}{Theorem}
\newtheorem*{lem*}{Lemma}
\newtheorem*{prop*}{Proposition}
\newtheorem*{cor*}{Corollary}
\newtheorem*{conj*}{Conjecture}

\theoremstyle{definition}
\newtheorem*{ass*}{Assumption}
\newtheorem*{dfn*}{Definition}

\theoremstyle{remark}
\newtheorem*{rem*}{Remark}

\def\dd{\mathrm{d}}
\def\ee{\mathrm{e}}

\newcommand*\circled[1]{\textcircled{\footnotesize#1}}

\usepackage{color}
\usepackage{hyperref}

\begin{document}

\title{\boldmath Spectral instability of parametrized black hole quasinormal modes\\
in the high-overtone limit via the exact WKB analysis}

\author{Taiga Miyachi}
\email{tmiyachi@omu.ac.jp}
\affiliation{Department of Physics, Osaka Metropolitan University, Osaka 558-8585, Japan}
\affiliation{Osaka Central Advanced Mathematical Institute (OCAMI),
Osaka Metropolitan University, 3-3-138 Sugimoto, Sumiyoshi, Osaka 558-8585, Japan}

\author{Ryo Namba}
\email{ryo.namba@riken.jp}
\affiliation{RIKEN Center for Interdisciplinary Theoretical and Mathematical Sciences (iTHEMS), Wako, Saitama 351-0198, Japan}
\affiliation{Research Center for the Early Universe (RESCEU), Graduate School of Science, The University of Tokyo, Tokyo 113-0033, Japan}

\author{Hidetoshi Omiya}
\email{omiya@tap.scphys.kyoto-u.ac.jp}
\affiliation{Department of Physics$,$ Kyoto University$,$ Kyoto 606-8502$,$ Japan}

\author{Naritaka~Oshita}
\email{naritaka.oshita@yukawa.kyoto-u.ac.jp}
\affiliation{Center for Gravitational Physics and Quantum Information,%\\ 
Yukawa Institute for Theoretical Physics, Kyoto University, 606-8502, Kyoto, Japan}
\affiliation{The Hakubi Center for Advanced Research, Kyoto University,\\
Yoshida Ushinomiyacho, Sakyo-ku, Kyoto 606-8501, Japan}
\affiliation{RIKEN Center for Interdisciplinary Theoretical and Mathematical Sciences (iTHEMS), Wako, Saitama 351-0198, Japan}

\preprint{YITP-25-188, KUNS-3084, OCU-PHYS-619, AP-GR-207}

\date{\today}

\begin{abstract}
We study the asymptotic behavior of parametrized black hole quasinormal modes (QNMs) in the high-overtone limit.
To gain insights into their analytical structure, we apply the exact WKB method, which was recently developed by the same authors. 
Our theoretical predictions are confirmed in good agreement with the numerical results based on Leaver's method.
For specific values of parametrization parameters that characterize deviations from general relativity, we find that the real part of asymptotic QNM frequencies diverges in the high-overtone limit, in sharp contrast to the case of a Schwarzschild black hole.
This demonstrates that the convergence of the real parts of high-overtone QNMs is a distinctive feature of general relativity, while parametrized corrections generically lead to divergent spectral behaviors.
\end{abstract}

\maketitle

\newpage
\tableofcontents

\newpage

\section{Introduction}

Black hole quasinormal modes (QNMs) are one of the important probes of gravity in the strong-field regime.
QNM frequencies, $\omega_{\ell m n}$, in the multipole modes of $(\ell, m)$ are unique to the mass and spin of the remnant black hole, where $n = 0, 1, 2,...$ is the overtone number.
They can be read from the ringdown following the merger phase of a binary black hole.

The QNM spectral instability or pseudospectrum\footnote{This does not mean an uncontrollable growth of the wave amplitude in the temporal domain, or $\mathrm{Im}\,\omega > 0$ in the convention used later in this paper.} is a characteristics that has attracted considerable attention with active discussions. 
Even with a tiny modification in the perturbation equation, e.g., in the Regge-Wheeler equation, the distribution of QNMs in the complex frequency plane is significantly destabilized \cite{Nollert:1996rf,Barausse:2014tra,Jaramillo:2020tuu,Jaramillo:2021tmt,Cheung:2021bol,Berti:2022xfj} (see also the recent review \cite{Berti:2025hly}).
Despite the spectral instability, it has turned out that the time-domain waveform is stable against a %tiny 
small modification in the perturbation equation, e.g., environment effects \cite{Barausse:2014tra,Destounis:2023ruj}.
Therefore, the QNM instability would not impact the ability of black hole spectroscopy.

Nevertheless, the strong instability of QNM overtones has been an interesting topic.
It is known that the real parts of QNM frequencies in the high-damping (or large-overtone) limit, $\displaystyle \lim_{n \to \infty} \text{Re}(\omega_{\ell m n})$, converge to specific values for several black hole solutions in general relativity (GR) \cite{Nollert:1993zz,Andersson:1995zk,Motl:2002hd,Motl:2003cd,Andersson:2003fh,Keshet:2007be,Keshet:2007nv}.
However, some subtleties arise between these values in certain cases. For example, $\text{Re}(\omega_{\ell m n})$ of the Reissner--Nordstr\"om black hole in the {\it Schwarszschild limit} appears to asymptote to $(8 \pi)^{-1} \log 5$ \cite{Andersson:2003fh,Cardoso:2024mrw}, whereas the Schwarzschild black hole has a different factor $(8 \pi)^{-1}\log 3$ in the high-overtone limit \cite{Hod:1998vk,Nollert:1993zz,Andersson:2003fh}.
It is not fully understood how this apparent discrepancy limits the robustness of the black hole spectroscopy.

The high-overtone convergence has often been discussed in the context of quantum gravity, where the asymptotic values have been interpreted as hints of underlying microscopic structures \cite{Hod:1998vk,Motl:2002hd,Andersson:2003fh,Keshet:2007nv,Maggiore:2007nq}. 
In \cite{Keshet:2007nv,Keshet:2007be}, a Bohr-Sommerfeld equation was derived, which leads to the uniform separation of $2 \pi T_{\rm H}$ in the high-overtone QNM frequencies.
Despite the interesting implications, a purely classical explanation for the origin of the convergence is not well understood.

To understand how general such an asymptotic behavior of QNMs in the high-overtone limit is, we explore the overtone distribution in the framework of the parametrized black hole perturbation equation \cite{Cardoso:2019mqo,McManus:2019ulj,Hirano:2024fgp,Ghosh:2023etd,Volkel:2022aca,Cano:2024jkd,Volkel:2022khh,Tang:2025qaq}, which provides a unified framework to comprehensively capture deviations from GR.
It offers an ideal setting to explore whether the convergence of $\text{Re}(\omega_{\ell m n})$ observed in GR is a universal property or an exceptional one.
In this work, we analyze the asymptotic parametrized QNMs using the framework of exact WKB analysis~\cite{Voros1983TheRO,10.1007/978-4-431-68170-0_1,AIF_1993__43_1_163_0,NIKOLAEV_2022,Nikolaev_2023,2024arXiv241017224N}\footnote{The applications to the black hole QNMs are formulated in \cite{Miyachi:2025ptm}.}. 
By introducing beyond-GR corrections to the Regge–Wheeler potential, we show that the asymptotic value of $\text{Re}(\omega_{\ell mn})$ in the limit of $n \to \infty$ can be divergent in the high-overtone limit in some special cases.
This indicates that the asymptotic convergence of $\text{Re}(\omega_{\ell mn})$, observed in GR, does not hold in general in the framework of the parametrized QNMs.

This paper is organized as follows. 
In Section \ref{sec:Parametrized_BH}, we review the framework of parametrized QNMs and describe how the WKB expansion parameter $\eta$ can be introduced.
We also demonstrate the derivation of the QNM condition, established in \cite{Miyachi:2025ptm}, in the case of the Schwarzschild black hole.
In Section \ref{sec:AsymQNMs}, we present our methodology, applying the exact WKB analysis to the analysis of the asymptotic parametrized overtones. 
Section \ref{sec:Conclusion} then summarizes our main results, discusses the instability of high overtones, and outlines potential future research directions, e.g., black hole physics and quantum gravity.
Technical derivations and details regarding our methodology are presented in the Appendices.
In this paper, we take the unit $c = G = 1$.

\section{Formulation}
\label{sec:Parametrized_BH}
In this section, we briefly review the parametrized QNMs and the exact WKB method, which will be employed in the subsequent computation of QNM frequencies.

\subsection{Parametrized QNM formalism}

Here, we summarize the parametrized QNM formalism, which comprehensively characterizes QNM frequencies with small theory-agnostic deformations from GR~\cite{Cardoso:2019mqo,McManus:2019ulj,Hirano:2024fgp}.
In the frequency domain, the $(\ell,m)$ component of the master variable, $\Psi(\omega, r)$, for the odd-parity sector of perturbations around the Schwarzschild background spacetime satisfies
\begin{align}\label{eq:masterEq}
    \left[f^2\frac{\dd^2}{\dd r^2}+f f'\frac{\dd}{\dd r}+\omega^2 - V_{\mathrm{RW}}(r)\right]\Psi(\omega,r) &= 0~, & f(r)&\coloneqq 1-\frac{1}{r}~, & f'&\coloneqq \frac{\dd f}{\dd r}\,,
\end{align}
where the dimensionful quantities are measured in the units of the Schwarzschild radius, i.e.~effectively taking $r_h\coloneqq2M=1$. 
$V_{\mathrm{RW}}$ is the Regge--Wheeler potential given by
\begin{align}
    V_{\text{RW}}:= f\left( \frac{L}{r^{2}} + \frac{1-s^2}{r^{3}}\right)~,
\end{align}
with $L= \ell(\ell+1)$.
The parametrized QNM formalism incorporates deviations from GR by replacing the potential barrier $V_{\rm RW}$ with $V_{\rm para}$, i.e.,
\begin{align}
    V_{\rm para} &\coloneqq V_{\rm RW} + \delta V(r)\,,\\
    \delta V(r) &\coloneqq f(r)\sum_{j=3}^{\infty}\alpha_j\, r^{-j}\,.
\end{align}
The constants $\alpha_j$ are dimensionless coefficients that parametrize deviations from GR\footnote{We do not consider $j \leq 2$ as those corrections would change the asymptotic behavior of the effective potential, i.e., $V_{\rm RW} \sim \ell (\ell + 1)/ r^2$, which would potentially indicate the modifications of 
%the asymptotic spacetime
the global structure of spacetime.}.
The task we should address in this formalism is therefore to obtain QNM frequencies from the following perturbation equation:
\begin{align}\label{eq:para_masterEq}
    \left[f^2\frac{\dd^2}{\dd r^2}+f f'\frac{\dd}{\dd r}+\omega^2 - V_{\mathrm{para}}(r)\right]\Psi(r) = 0\,.
\end{align}
QNMs are the eigenvalues of $\omega$ for which the homogeneous solution to the master equation~\eqref{eq:para_masterEq} simultaneously satisfies the ingoing boundary condition at the horizon ($r = 1$)\footnote{
In general, deviations from the Schwarzschild spacetime modify the horizon structure. However, we assume that the horizon structure remains the same as that of the Schwarzschild spacetime. 
For more general scenarios in which this is not the case, the modification in the tortoise coordinate propagates to correction terms in the potential barrier (for details, see Appendix in Ref. \cite{Ghosh:2023etd}).
} and outgoing boundary condition at infinity ($r  = +\infty$), that is
\begin{align}
    \Psi \longrightarrow
    \begin{cases}
        \ee^{- i \omega r_*}~, &  (r \to 1)~,\\
        \ee^{+ i \omega r_*}~, & (r \to +\infty)\,.
    \end{cases}
    \label{eq:boundary_condition}
\end{align}
Here, $r_*$ is the tortoise coordinate given by $r_* = r + \log|r -1|$. There exist several methods to compute QNM frequencies: the direct integration \cite{Chandrasekhar:1975zza}, Leaver's method~\cite{Leaver:1985ax}, WKB method~\cite{Iyer:1986np}, monodoromy~\cite{Andersson:2003fh}, spectral method \cite{Chung:2023zdq}, and exact WKB method~\cite{Miyachi:2025ptm} that was recently %developed 
applied to the QNMs of Schwarzschild black holes by the authors. 
In the following, we apply the exact WKB method to the parametrized black hole perturbations to explore the asymptotic behavior of their QNM frequencies. 

\subsection{Exact WKB method}

We here briefly review the exact WKB method based on ~\cite{kawai2005algebraic,Miyachi:2025ptm,Namba:2025ejw}.
To obtain the WKB solutions, we first recast the master equation Eq.~\eqref{eq:para_masterEq} into a Schr\"{o}dinger-type equation. 
With the transfromation $\psi \coloneqq f^{1/2} \Psi$, we obtain
\begin{align}
	\bigg[\frac{\dd^{2}}{\dd r^{2}} + \bigg(\frac{\omega^{2}-V_{\text{para}}}{f^{2}} + \frac{f^{\prime 2}}{4f^{2}} - \frac{f''}{2f} \bigg)  \bigg]\psi = 0~.
    \label{eq:mastereq_pBH}
\end{align}
Furthermore, we insert a formally large parameter $\eta$ that admits the formal solutions in terms of the WKB series expansion as
\begin{align}
	\bigg[-\frac{\dd^{2}}{\dd r^{2}} + \eta^{2}Q_{\text{para}}(r,\eta) \bigg]\psi = 0\,,
    \label{eq:SchEq_eta}
\end{align}
where
\begin{subequations}
\begin{align}
    &Q_{\text{para}}(r,\eta) = Q_{\text{RW},0}(r) + \delta Q(r) + \eta^{-2}Q_{\text{RW},2}(r) 
    ~, \label{eq:Qpara3}\\
    &Q_{\text{RW},0}(r) = \frac{1}{r^{2}(r-1)^{2}} \left[ -\omega^{2}r^{4}+ \left( r-1 \right) \left( Lr-s^2 \right) \right] \label{eq:Q0}~,\\
    &Q_{\text{RW},2}(r) =  -\frac{1/4}{r^{2}(r-1)^{2}}~,\\
    &\delta Q(r) = \sum_{j=3}^{\infty} \frac{\alpha_j}{r^{j-1}(r-1)}
    \eqqcolon \sum_{j=3}^{\infty} \delta Q_j(r)\,.
    \label{eq:deltaQ_para}
\end{align}
\end{subequations}
This reduces to the original equation \eqref{eq:SchEq_eta} with $\eta = 1$, which makes sense after the resummation of the full WKB series performed below.
We then solve the equation order by order in terms of $\eta$ with the ansatz
\begin{align}
    \psi_{\pm}(r,\eta) &=  \exp\bigg(\pm \int_{r_i}^r S(r')\,\dd r'\bigg)~, \qquad S \coloneqq \sum_{j=-1}^{\infty}\eta^{-j} S_{j} \ ,
\end{align}
where the would-be arbitrary integration endpoint $r_i$ is taken to be at a turning point of $Q_0$, i.e.~$Q_0(r_i) = 0$, for the sake of the convenience for exact WKB formulation.
The functions $S_j$ satisfy the recursion relation
\begin{subequations}
\label{eq:recursion}
\begin{align}
    &S_{-1}^2=Q_{0}(r) = Q_{\text{RW},0}(r) + \delta Q(r)  \; , \qquad
    S_0 = - \frac{1}{2 S_{-1}}\frac{\dd S_{-1}}{\dd r} \; , \\
    & S_{1} = - \frac{1}{2 S_{-1}} \left( \frac{\dd S_0}{\dd x} + \;  S_0^2 - Q_{\text{RW},2} \right) \; ,\\
    & S_{j+1} = - \frac{1}{2 S_{-1}} \left( \frac{\dd S_j}{\dd x} + \; \sum_{\mathclap{\substack{n+m = j \\ n \ge 0 , m \ge 0}}} S_n S_m \right) \; , \quad 
    \left( j= 1,2, \dots \right) \; .
    \label{eq:S_j}
\end{align}
\end{subequations}
The resulting WKB solutions are given by
\begin{align}
    \psi_{\pm}(r,\eta) &= \frac{1}{\sqrt{S_{\text{odd}}}}\exp\bigg(\pm \int_{r_i}^r S_{\text{odd}}(r')\,\dd r'\bigg)~, \qquad  S_{\text{odd}} \coloneqq\sum_{j=0}^{\infty}\eta^{1-2j} S_{2j-1}~.
    \label{eq:WKB_solutions}
\end{align}
The $\pm$ sign in Eq.~\eqref{eq:WKB_solutions} corresponds to the branch choice when solving $S_{-1}^2=Q_{0}$.

The choice of $\eta$ insertion for $Q_{\text{para}}$ is based on the previous paper~\cite{Miyachi:2025ptm}.
Under the insertion, the asymptotic behaviors of the leading-order WKB solutions $\psi^{\rm WKB}_\pm \sim Q_{0}^{-1/4} \exp \left( \pm \int^r \sqrt{Q_{0}} \, \dd r \right)$ around the singular points $r = 1$ and $\infty$ (with $\eta=1$) become%
\footnote{In our previous paper \cite{Miyachi:2025ptm}, the asymptotic behavior around $r=0$ is correctly reproduced once those around $r=1,\infty$ are imposed.
The parametrized black hole under consideration has additional freedom associated with the correction $\delta Q$, and the behavior around $r=0$ depends on it. 
We here follow the choice of $\eta$ insertion in \cite{Miyachi:2025ptm} for simplicity as well as to make the comparison with our previous results straightforward. 
}
\begin{align}
\label{eq:psiWKB_limits}
    \psi_{\pm} \sim
    \begin{cases}
        \displaystyle
        (r-1)^{1/2 \pm i\omega} \; , \qquad 
        & r\to 1 \; , \vspace{1mm}\\
        \displaystyle
        \ee^{\pm i\omega r} r^{\pm i\omega} \; , \qquad
        & r\to \infty \; ,
    \end{cases}
\end{align}
where we take the branch of $\sqrt{Q_{0}}$ as
\begin{align}
    \sqrt{Q_{0}}
    \simeq
    \begin{cases}
        \displaystyle
        \frac{i\omega}{r-1}, \quad & r\to1, \vspace{1mm}\\
        \displaystyle
        \left(1+\frac{1}{r}\right)i\omega, \quad & r\to\infty.
    \end{cases}
\end{align}
Then, \eqref{eq:psiWKB_limits} matches with the asymptotic solutions to the original differential equation \eqref{eq:mastereq_pBH}.
To identify the QNM frequency $\omega$, we need to impose proper boundary conditions \eqref{eq:boundary_condition}.
In terms of WKB solutions, these conditions correspond to associating
\begin{align}
\psi \sim
\begin{cases}
\psi_{-} \ , \qquad & \mbox{in the vicinity of } r= 1 \,,\\
\psi_{+} \ , \qquad & \mbox{in the vicinity of } r= \infty \,,
\end{cases}
\label{eq:QNMcondition}
\end{align}
according to Eqs.~\eqref{eq:psiWKB_limits}.

This association should be understood as in the asymptotic sense, and the exact identification could be done only if the WKB series converges.
In general, however, the WKB series obtained as an asymptotic expansion in the above manner is divergent.
The key idea of the exact WKB analysis is to apply the Borel resummation to this divergent series in order to obtain well-defined solutions \cite{Voros1983TheRO}.
Formally, expanding the WKB solutions \eqref{eq:WKB_solutions} by $\eta$, we obtain
\begin{align}
    \psi_{\pm}(r,\eta)=\ee^{\pm\eta\,y_0(r)}\sum_{n=0}^{\infty}\eta^{-n-1/2}f_n(r), 
\end{align}
where $y_0(r)\coloneqq \int^r_{r_i}\sqrt{Q_0(r')}\dd r'$ and $f_n(r)$ are complex functions.
Its Borel transform is defined by
\begin{align}
    \mathcal{B}[\psi_{\pm}](r,y)=\sum_{n=0}^{\infty} \frac{f_n(r)}{\Gamma(n+1/2)}[y\pm y_0(r)]^{n-1/2},
    \label{eq:Borel_transf}
\end{align}
where $\Gamma(x)$ is the gamma function, and the corresponding Borel sum is the Laplace transform of $\mathcal{B}[\psi_{\pm}]$,
\begin{align}
    \mathcal{S}[\psi_{\pm}](\eta,r)=\int_{\mp y_0}^{\infty} \ee^{-\eta y}\mathcal{B}[\psi_{\pm}](r,y)\,\dd y,
    \label{eq:BorelSum}
\end{align}
with the integration contour is $y\in \{\mp y_0(r)+t;, t\geq0\}$ and positive $\eta$ unless otherwise noted.

If the series $\sum \eta^{-n} f_n$ falls into the Gevrey-$1$ type, if $\mathcal{B}[\psi_{\pm}]$ is analytically continued in the region that contains the integration path, and if the Laplace integral in Eq.~\eqref{eq:BorelSum} yields a finite and definite value for sufficiently large $\eta$, then the original WKB solutions are said to be {\it Borel summable}.
The Borel summability depends on the variable $r$, and it is known that the Borel summability may not hold on {\it the Stokes curves}, which are defined on the complex plane by the following criterion:
\begin{align}
    \mathrm{Im}\, y_0(x) = \mathrm{Im}\,\int^r_{r_i}\sqrt{Q_{0}(r')}\dd r' =0.
\end{align}

By analyzing the well-defined solution obtained from the Borel sum, one can give a rigorous description of the Stokes phenomenon, which in turn allows a global characterization of the solution. 
In what follows, we present a relevant assumption and theorem; see Refs.~\cite{kawai2005algebraic,2014arXiv1401.7094I,Miyachi:2025ptm,Namba:2025ejw} for detailed descriptions.

\begin{figure}
    \centering
    \includegraphics[width=0.5\linewidth]{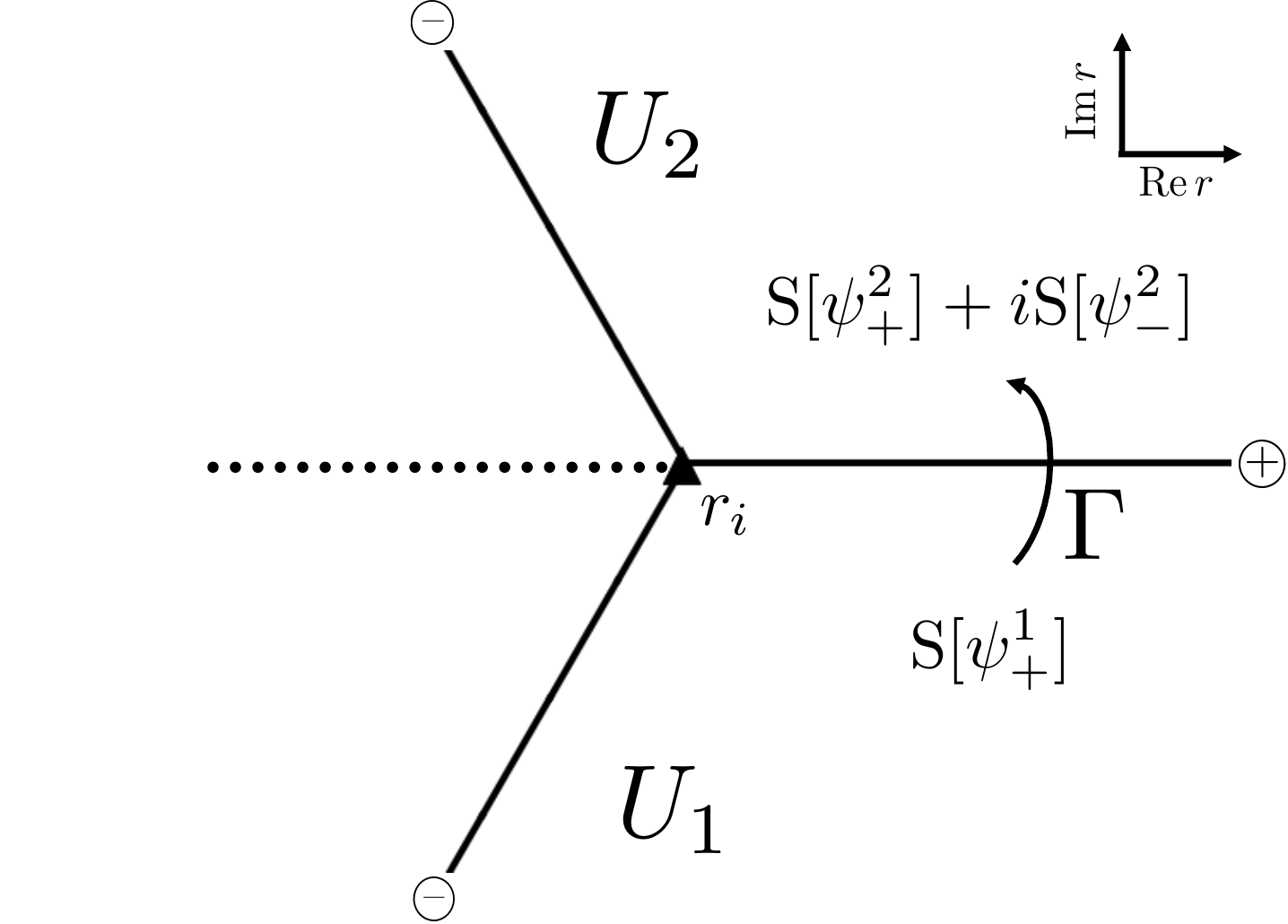}
    \caption{Schematic picture of the Stokes curves for the potential $Q_{0}(r)=r-r_i$ showing the branch cut and the counter-clockwise analytic continuation from the Stokes region $U_1$ to $U_2$ across the Stokes curve $\Gamma$. 
    The triangle marker represents a turning point at $r = r_i$.
    The branch cut originating from $\sqrt{Q_0(r)}$ is shown as the dotted line. The dominant WKB solution, $\mathcal{S}[\psi_{+}]$ or $\mathcal{S}[\psi_{-}]$, on each Stokes curve is denoted by \circled{+} or \circled{-}, respectively. We use these symbols throughout this paper.}
    \label{fig:SL_r-r_i}
\end{figure}

First, we impose the following assumption on $Q_{0}$:
\begin{ass}
    $Q_{0}$ is a holomorphic function, that all turning points (zeros) of $Q_{0}$ are simple, and that all poles of $Q_{0}$ are of order two or higher.
    We also assume that no two turning points are connected by Stokes curves.
    \label{ass:Q0}
\end{ass}
Under Assumption~\ref{ass:Q0}, the following theorem holds for the Stokes curves.
\begin{thm}
    \label{thm:stokes}
    Suppose two Stokes regions $U_1$ and $U_2$ that have a Stokes curve $\Gamma$ as a common boundary, and this curve emerges from a turning point $r_i$ (see Figure~\ref{fig:SL_r-r_i}).
    Let $\mathcal{S}[\psi_{\pm}^{i}]\,(i=1,2)$ be the Borel sums in $U_{i}\,(i=1,2)$. Then, the Borel sums $\mathcal{S}[\psi_{\pm}^{i}]$ are analytically continued from $U_1$ to $U_2$ as follows:
    \begin{align}
        \begin{cases}
            \displaystyle
            \mathcal{S}[\psi_{+}^{1}] = \mathcal{S}[\psi_{+}^{2}] \pm i\mathcal{S}[\psi_{-}^{2}],&\quad 
            \displaystyle
            \mathcal{S}[\psi_{-}^{1}] = \mathcal{S}[\psi_{-}^{2}]
            ,\quad \left( \mathrm{Re}\int_{r_i}^r\sqrt{Q_{0}(r')}\,\dd r'>0 \right) \vspace{2mm}\\
            \displaystyle
            \mathcal{S}[\psi_{-}^{1}] = \mathcal{S}[\psi_{-}^{2}]\pm i\mathcal{S}[\psi_{+}^{2}]
            ,&\quad 
            \displaystyle
            \mathcal{S}[\psi_{+}^{1}] = \mathcal{S}[\psi_{+}^{2}],\quad \left( \mathrm{Re}\int_{r_i}^r\sqrt{Q_{0}(r')}\,\dd r'<0 \right)
        \end{cases}
        \label{eq:analytic_continuation}
    \end{align}
    The signs $+$ and $-$ correspond to the counter-clockwise and clockwise crossing of $\Gamma$, respectively, from $U_1$ to $U_2$ with respect to the turning point $r_i$.
    When $\mathrm{Re}\int_{r_i}^r\sqrt{Q_{0}(r')}\,\dd r'>0$, we call $\mathcal{S}[\psi_{+}]$ is dominant over $\mathcal{S}[\psi_{-}]$ and vice versa. 
\end{thm}
Note that for the potentials we consider in this paper, Assumption~\ref{ass:Q0} is satisfied, and thus Theorem~\ref{thm:stokes} follows.
Figure~\ref{fig:SL_r-r_i} is a schematic picture showing the Stokes curves and associated Stokes phenomenon for the simplest case of $Q_0(r)=r-r_i$.
Under Assumption \ref{ass:Q0}, three Stokes curves emanate from a simple turning point $r_i$, and each flows toward either infinity or a pole of $Q_0(r)$.

Not only (i) the Stokes phenomenon but also (ii) the switching of turning points when multiple Stokes curves contribute, as well as (iii) crossing branch cuts associated with $1/\sqrt{S_{\rm odd}}$ in \eqref{eq:WKB_solutions}, are involved in the analytic continuation of the WKB solutions.
It is convenient to represent the analytic continuation \eqref{eq:analytic_continuation} by $2\times2$ matrix $M$ as~\footnote{We can omit the Borel sum operator $\mathcal{S}$ because $\mathcal{S}[f+g]=\mathcal{S}[f]+\mathcal{S}[g]$.}
\begin{align}
    \label{eq:matrix_analytic_continuation}
    \begin{pmatrix}
        \psi_{+}\\
        \psi_{-}
    \end{pmatrix}^{1}
    =M
    \begin{pmatrix}
        \psi_{+}\\
        \psi_{-}
    \end{pmatrix}^{2}\,,
\end{align}
where the superscripts $1$ and $2$ indicate the label of the corresponding Stokes regions.
The building blocks of the matrix $M$, consisting of (i)-(iii), are as follows.%
\footnote{We restrict ourselves to being concise in this review part. For a detailed description of the usage of the effects due to (i)-(iii), we refer interested readers to \cite{Miyachi:2025ptm} and the references therein.}

\subsubsection*{(i) Stokes phenomena}
When the analytic continuation is performed by a path crossing a Stokes curve $\Gamma$ counterclockwise around its associated turning point, the analytic continuation can be written in the $2\times 2$ matrix form as
\begin{align}
\label{eq:connection_matrix}
    \begin{pmatrix}
        \psi_+\\
        \psi_-
    \end{pmatrix}^{1}
    = V_{\pm}
    \begin{pmatrix}
        \psi_+\\
        \psi_-
    \end{pmatrix}^{2}~,
\end{align}
with
\begin{align}\label{eq:connectionS}
            V_+ =
            \begin{pmatrix}
                1 & i\\
                0 & 1
            \end{pmatrix}~, 
            \qquad 
            V_- =
            \begin{pmatrix}
                1 & 0\\
                i & 1
            \end{pmatrix}~.
    \end{align}
The $+$ ($-$) subscript corresponds to the case when the solution $\psi_{+}$ ($\psi_{-}$) is dominant on the Stokes curve $\Gamma$.
For a clockwise crossing, the sign in front of $i$ is flipped, that is, $V_\pm^{-1}$ should be used in the place of $V_\pm$ in \eqref{eq:connection_matrix}.
This change of coefficients reflects the effect of the Stokes phenomenon.

\subsubsection*{(ii) Changing turning points}
In general, multiple turning points exist, requiring us to cross multiple Stokes curves originating from different turning points. 
In such a situation, one needs to change the reference turning points of the integral in Eq.~\eqref{eq:WKB_solutions} as one encounters a new Stokes curve emanating from another turning point.
Changing the reference turning point from $r_i$ to $r_j$, we multiply the matrix
\begin{align}\label{eq:connectionT}
    \Gamma_{ij} =
    \begin{pmatrix}
        u_{ij} & 0\\
        0 & u_{ij}^{-1}
    \end{pmatrix}~, \qquad
    u_{ij} = \exp\bigg(\int_{r_i}^{r_j} S_{\text{odd}}\, \dd r\bigg)\,.
\end{align}
in the manner similar to \eqref{eq:connection_matrix}.

\subsubsection*{(iii) Branch cuts from regular singular points}
The WKB solutions also acquire a nontrivial phase factor when one needs to cross a branch cut emanating from a singular point in $Q_0(x)$, which is associated with the factor $1/\sqrt{S_{\rm odd}}$ in the WKB solution \eqref{eq:WKB_solutions}. 
In the case of regular singular points, i.e.~$Q_0 \propto (x- b_i)^{-2}$ around the $i$-th singular point $b_i$, the factor $1/\sqrt{S_{\rm odd}} \sim (\eta^2 Q_0)^{-1/4}$ is two-valued around $b_i$, while the exponential factor is single-valued. 
Therefore, when $\psi_{\pm}$ crosses a branch cut associated with a regular singular point, we multiply the matrix 
\begin{align}
    D_i &= 
    \begin{pmatrix}
        \nu_i^+ & 0\\
        0 & \nu_i^-
    \end{pmatrix}~, \quad
    \nu_i^{\pm} = \exp\left[ i\pi \, \big( 1\pm 2 \underset{x=b_i}{\text{Res}}\,S_\text{odd} \big) \right] \ ,
    \label{eq:MatrixD}
\end{align}
in the manner similar to \eqref{eq:connection_matrix}.
In the case of the potential~\eqref{eq:SchEq_eta}, $\nu_i^{\pm}$ becomes~\cite{Miyachi:2025ptm}
\begin{align}
    \nu_1^{\pm} 
    = \exp\left[ i\pi \, \big( 1\pm 2 \underset{r=1}{\text{Res}}\,S_\text{odd} \big) \right]
    = - \ee^{\mp2\pi\omega\eta} \ ,
    \label{eq:nu_1}
\end{align}
for the branch cut emerging from the event horizon $r=1$.

\subsection{Applications to black hole QNMs}

\begin{figure}
    \centering
    \includegraphics[width=0.8\linewidth]{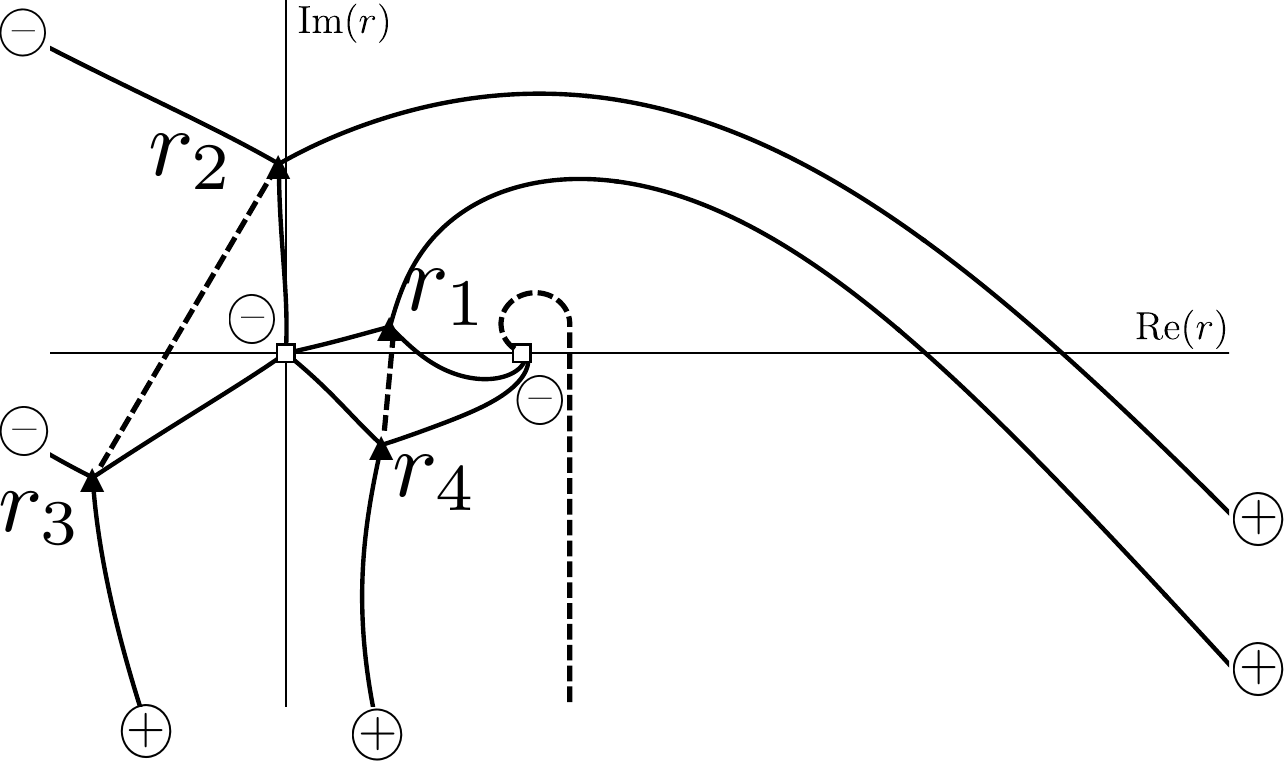}
    \caption{Schematic picture of the Stokes curves for the potential $Q_{\text{RW},0}$ with the branch cuts (see Fig.~\ref{fig:SL_r-r_i} for the explanation of the symbols). The open squares represent the poles of $Q_0(r)$, $r=0$ and $r=1$. For illustration, we take $\ell=s=2$ and $\omega = 3.0 - 3.3 i$. The curves flowing to $r = 1$ forms a logarithmic spiral.}
    \label{fig:SL_Schwarzschild}
\end{figure}

To illustrate the application of the exact WKB analysis to black hole QNMs, we consider the case of the Schwarzschild black hole $Q_0=Q_{\mathrm{RW},0}$ based on \cite{Miyachi:2025ptm}.
Since we focus on the eigenvalues of higher overtones, we assume large $\mathrm{Im}\,\omega$ throughout this paper. The Stokes curves of the Schwartzschild case~$Q_0=Q_{\mathrm{RW},0}$ are depicted in Fig.~\ref{fig:SL_Schwarzschild}. Note that the turning points, $r_i$, are determined by zeros of the potential~\eqref{eq:Q0}:
\begin{align}
    -\omega^2 r_i^4 + (r_i-1) (L r_i - s^2) = 0\,.
\end{align}
In the case of large $|\omega|$, the four turning points are at $r^4 \sim s^2/\omega^2$, forming an approximate square around the origin.

To obtain the QNM frequencies, one has to impose the boundary conditions \eqref{eq:QNMcondition} on the WKB solution.
Let us consider a WKB solution $\psi_-$ near the horizon and $\psi_+$ at the far zone.
The two solutions are related via the analytic continuation determined by the configuration of the Stokes curves, turning points, and branch cuts over the complex-$r$ plane, which can be incorporated into a single $2 \times 2$ matrix $M$.
Therefore, the QNM condition can be obtained by
\begin{equation}
    M_{22} (\omega) = 0\,,
    \label{eq:m22qnmcondition}
\end{equation}
where $M_{21}(\omega) \neq 0$ should also hold.
In the following, we briefly review the derivation of the QNM condition $M_{22} (\omega) = 0$ \cite{Miyachi:2025ptm}.

In \cite{Miyachi:2025ptm}, we pointed out the existence of a logarithmic spiral and a branch cut from the regular singular point $r=1$. 
Thus, when analytically continuing the WKB solutions along the real axis from $r = 1$ to $r = +\infty$, the contour must cross the two distinct Stokes curves, one connecting $r = r_1$ and $r = 1$, and the other connecting $r = r_4$ and $r = 1$, as well as the branch cut an infinite number of times, see Fig.~\ref{fig:SL_Schwarzschild}. 
Furthermore, it also crosses two additional Stokes curves connected to $r=+\infty$, one emanating from $r = r_1$ and the other from $r = r_2$.
Taking into account these contributions, we can calculate the analytic continuation from $r=1$ to $r=\infty$ by using \eqref{eq:connectionS}, \eqref{eq:connectionT} and \eqref{eq:MatrixD}. 
The resulting connection matrix is given by
\begin{align}
          M & = \lim_{n\to\infty} \left[ \left( V_- \right)^{-1} \Gamma_{14} \left( V_- \right)^{-1} \Gamma_{41} \, D_1 \right]^n V_+ \Gamma_{12} V_+ \Gamma_{21} 
          \nonumber\\
        &=\lim_{n\to\infty}
          \begin{pmatrix}
            \displaystyle
            \left( \nu^+_1 \right)^n
            & \displaystyle
              i \left( \nu^+_1 \right)^n \left( 1 + u_{12}^2 \right) \vspace{2mm} \\
            \displaystyle
            -i \nu_1^+ \, \frac{\left( \nu_1^- \right)^n - \left( \nu_1^+ \right)^n}{\nu_1^- - \nu_1^+}(1 + u_{41}^2) \quad
            & \displaystyle
              \left( \nu^-_1 \right)^n + \nu_1^+ \, \frac{\left( \nu_1^- \right)^n - \left( \nu_1^+ \right)^n}{\nu_1^- - \nu_1^+} \left( 1 + u_{41}^2 \right) \left( 1 + u_{12}^2 \right)
          \end{pmatrix}\,.
          \label{eq:contourmatrix_Schwarzschild}
\end{align}
Then the QNM condition \eqref{eq:m22qnmcondition} reduces to

\begin{align}
\lim_{n\to\infty}
    (\nu_1^{-})^n \left[ 1+ \nu_1^+ \, \frac{1 - \left( \frac{\nu_1^+}{\nu_1^-} \right)^n}{\nu_1^- - \nu_1^+} \left( 1+u_{41}^2 \right) \left( 1+u_{12}^2 \right) \right]=0\,.
    \label{eq:M22_Schwarzschild}
\end{align}
Using \eqref{eq:nu_1} and our implicit condition ${\rm Re} \left( \omega \right) > 0$, we have $\mathrm{Re} \left( \nu_1^+/\nu_1^- \right)=\mathrm{Re}\,\ee^{-4\pi\omega} < 1$ (with $\eta=1$) and consequently $\big( \nu_1^+ / \nu_1^- \big)^n \to 0$ in the limit $n\to\infty$.
Then the condition \eqref{eq:M22_Schwarzschild} reduces to
\begin{align}
\label{eq:QNMcondition_deltaQ3}
    \ee^{4\pi\omega} + u_{12}^2+u_{41}^2+u_{12}^2u_{41}^2=0\,.
\end{align}
It is convenient to rewrite this condition as
\begin{align}
    \omega = \frac{1}{4\pi}\log(u_{12}^2+u_{41}^2+u_{12}^2u_{41}^2) 
    - \frac{i}{2} \left( N + \frac{1}{2} \right),
    \label{eq:QNMcondition_re}
\end{align}
where $N$ is a large positive integer.

In Appendix \ref{app:phase_int_Q3}, we perturbatively calculate the relevant phase integrals $u_{12}$ and $u_{41}$ for large $\omega$. The results are
\begin{align}
    u_{12}^2 &= \ee^{i\pi s} \left[ 1+2c_2 \left( 1-i \right) \omega^{-1/2}+\mathcal{O} \left( \omega^{-3/2} \right) \right]~,\\
    u_{41}^2 &= \ee^{-i\pi s} \left[ 1+2c_2 \left( 1+i \right) \omega^{-1/2}+\mathcal{O} \left( \omega^{-3/2} \right) \right]~,\\
    c_2 &= \frac{\Gamma(1/4)^2}{24\sqrt{2\pi}s^{1/2}} \left( 3L-s^2 \right)~.
\end{align}
Substituting into the condition~\eqref{eq:QNMcondition_re} and solving iteratively with respect to the perturbation coefficient $(N + 1/2)^{-1}$, we obtain
\begin{align}
    \omega \simeq \frac{1}{4\pi} \log\left[ 1+2\cos \pi s + \gamma {\left( N + \frac{1}{2} \right)^{-1/2} }\right]
    {- \frac{i}{2} \left( N + \frac{1}{2} \right)}~,
    \label{eq:QNM_analytic}
\end{align}
where 
\begin{align}
    \gamma = \frac{\ee^{i\pi/4}\Gamma(1/4)^2}{6\sqrt{\pi s}} \left[ 1+\sqrt{2} \cos\pi \bigg( s - \frac{1}{4} \bigg) \right] \left( 3L-s^2 \right)~.
    \label{eq:gamma_coefficient}
\end{align}
In the limit $N \to \infty$, the real part of $\omega$ takes the well-known value $\mathrm{Re}\,\omega \simeq \log(1+2\cos\pi s)/(4\pi)$ ~\cite{Hod:1998vk,Motl:2002hd,Motl:2003cd,Andersson:2003fh}.

Let us emphasize that the QNM condition \eqref{eq:QNMcondition_deltaQ3}, or \eqref{eq:QNMcondition_re}, is exact and applicable independently of values of the model parameters, as long as the Stokes structure stays the same. Approximations are needed only at the last stage of calculation, to compute the phase factors $u_{ij}$ due to the changes of turning points. If the Stokes structure differs for a different parameter regime, and if different Stokes curves and/or branch cuts lie along the path of analytic continuation, one has to re-compute the matrix $M$ in \eqref{eq:matrix_analytic_continuation} following the rules described above.

\section{Asymptotic behavior of QNMs in parametrized black holes}
\label{sec:AsymQNMs}

This section is devoted to deriving the main result of this paper, by applying the exact WKB analysis to the perturbation equation of the parametrized black hole with the corrections $\delta Q_3$ and $\delta Q_4$, defined in \eqref{eq:deltaQ_para}.
Also, we discuss the case of $\delta Q_{p\geq5}$ and of general corrections.

\subsection{$\delta Q_3$ correction}
\label{subsec:deltaQ3}
Let us first consider the correction $\delta Q_3$.
The potential is given by
\begin{align}
    Q_{0}^{(3)}\coloneq
    Q_{\text{RW,0}}+\delta Q_3 = \frac{-\omega^2r^4+(r-1)(Lr-s^2+\alpha_3)}{r^2(r-1)^2}~.
    \label{eq:Q_para0^3}
\end{align}
It can be seen that the correction is only a replacement $s^2 \to s^2 - \alpha_3$ for the Regge-Wheeler potential \eqref{eq:Q0}.%
\footnote{In this paper, we assume $\alpha_3<s^2$ because the Stokes geometry is the same as the Regge-Wheeler equation with spin $s=1,2$ (Fig.~\ref{fig:SL_Schwarzschild}).}
Hence the higher overtones are obtained simply by replacing $s^2 \to s^2 - \alpha_3$ in the final result \eqref{eq:QNM_analytic} and \eqref{eq:gamma_coefficient}.
We thus find that the real part of the QNM frequencies converges to $(4\pi)^{-1}\log(1 + 2 \cos(\pi \sqrt{s^2  - \alpha_3}))$ for high overtones.
This result is applicable for arbitrarily small values of $\alpha_3$, which is in a more physically relevant regime, and recovers the Schwarzschild case in the limit $\alpha_3 \to 0$.

On the other hand, 
when $\alpha_3$ satisfies $1+2\cos(\pi\sqrt{s^2-\alpha_3})=0$, i.e.,
\begin{align}
    \alpha_3
    =s^2-4\bigg(k\pm\frac{1}{3}\bigg)^2, \quad (k\in \mathbb{Z})~,
    \label{eq:div_alpha_3}
\end{align}
the QNM frequencies are
\begin{align}
    \omega \simeq \frac{1}{4\pi}\log\left[ \gamma_3 {\left( N + \frac{1}{2} \right )^{-1/2}}\right] 
    {- \frac{i}{2} \left( N + \frac{1}{2} \right)}~,
    \label{eq:qnmlogNdiv}
\end{align}
where $\gamma_3$ is the constant obtained from \eqref{eq:gamma_coefficient} with the replacement of $s^2\to s^2-\alpha_3$.
This indicates that the real part of QNMs diverges as $\text{Re}(\omega) \propto \log N$ in the high-overtone limit for the specific values of $\alpha_3$ in Eq.~\eqref{eq:div_alpha_3}. 
This result may affect the conjecture on the relation between high-overtone QNMs and the quantum gravitational picture \cite{Hod:1998vk} as the real part of QNM frequencies represents the energy of quanta absorbed by a black hole.
The divergence of $\text{Re}(\omega)$ in $N \to \infty$ is a non-trivial result, as setting $\alpha_3$ to the specific values does not alter the configuration of the turning points, Stokes curves, or branch cuts.

To cross-check the prediction of the exact WKB analysis, we numerically calculate the QNM frequencies with large $N$ based on Leaver's method (see Appendix~\ref{app:C} for details). In Fig.~\ref{fig:Rew_j3_alpha_div}, we show the results for the case of $\ell=s=2$ with $\alpha_3=20/9$ and $32/9$, for which the $\log$-$N$ divergence of the real part is predicted as in \eqref{eq:qnmlogNdiv}.  
We confirm the excellent agreement between the numerical computation (Leaver's method) and our analytic prediction, based on the exact WKB analysis, in the large $N$ regime.

\begin{figure}
    \centering
    \begin{tabular}{cc}   
        \begin{minipage}[b]{0.5\linewidth}
            \centering
            \includegraphics[keepaspectratio, scale=0.33]{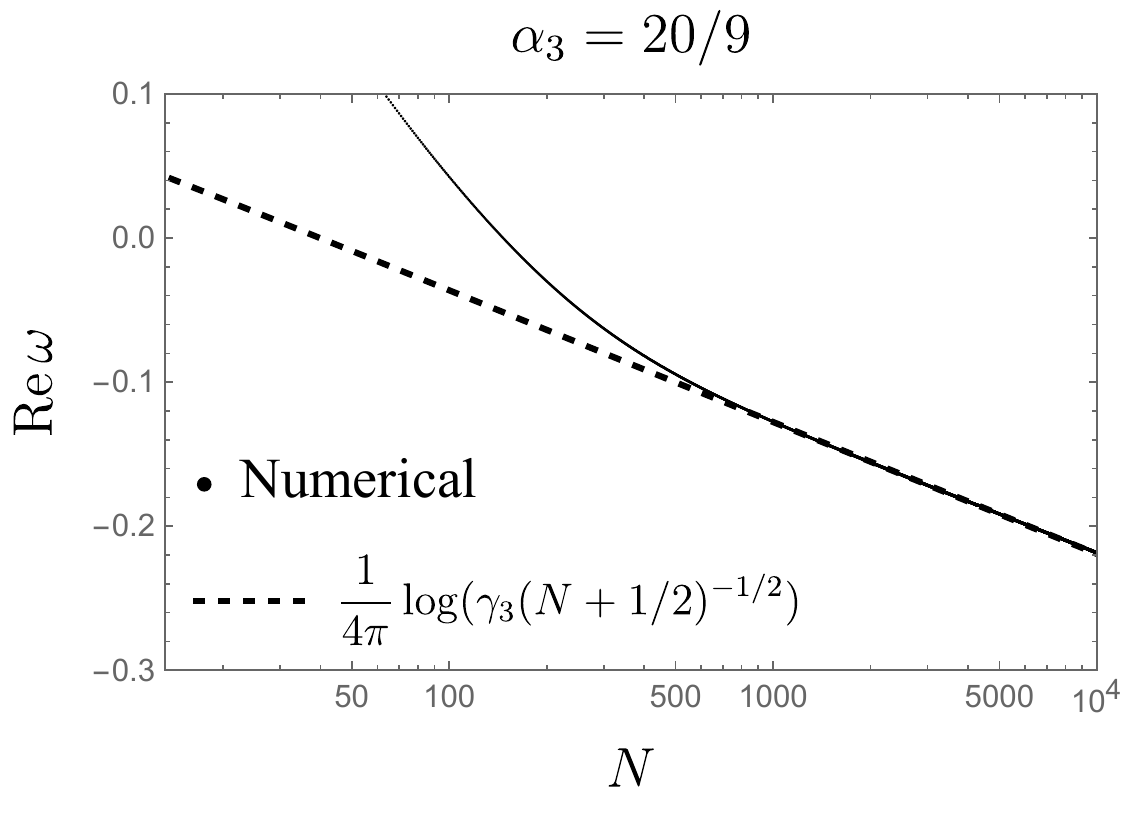}
        \end{minipage}&
        \begin{minipage}[b]{0.5\linewidth}
                \centering
                \includegraphics[keepaspectratio, scale=0.33]{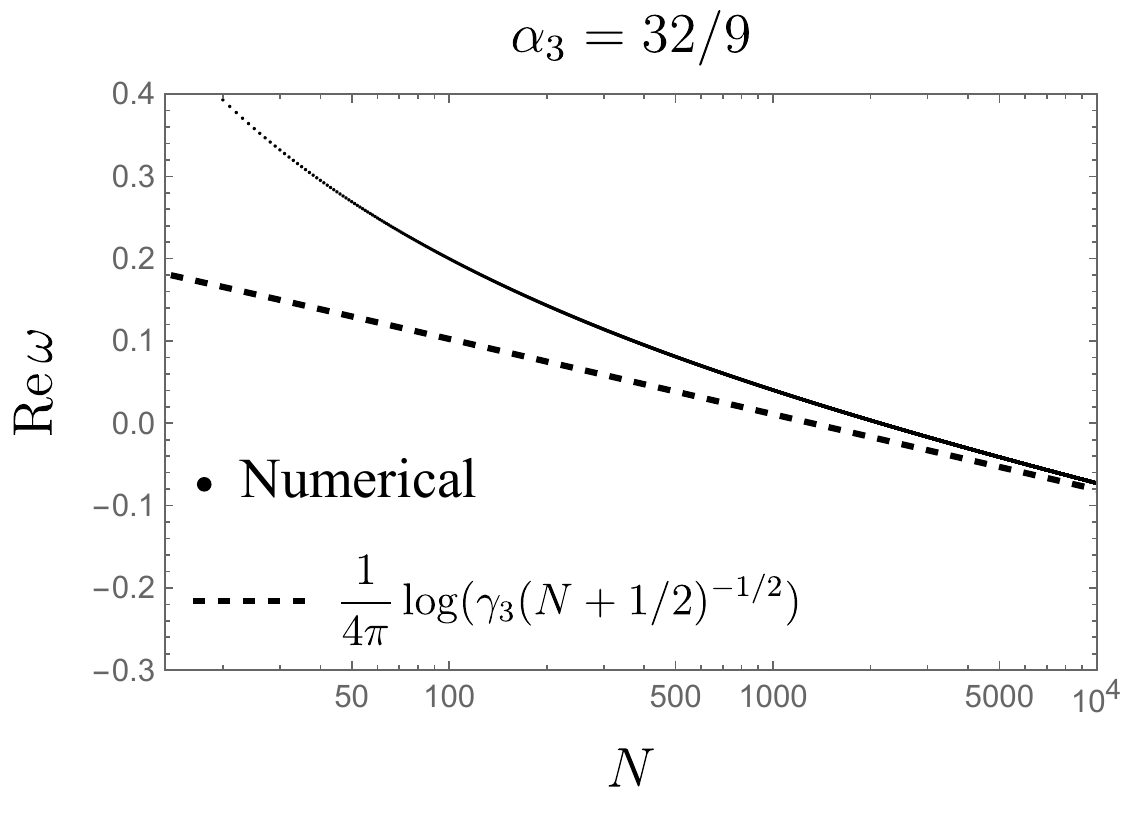}
        \end{minipage}
    \end{tabular}
    \caption{The asymptotic behavior of $\text{Re}(\omega_N)$ for the case of $\delta Q_3$. The black (dense) dots and black dashed line correspond to the numerical result (Leaver) and theoretical prediction (exact WKB), respectively. We take $\alpha_3 = 20/9$ (left) and $\alpha_3 = 32/9$ (right) with $\ell = s = 2$, for which $\text{Re}(\omega_N) \to \infty$ in the high-overtone limit $N \to \infty$.}
    \label{fig:Rew_j3_alpha_div}
\end{figure}

\subsection{$\delta Q_4$ correction}
\label{subsec:deltaQ4}
\begin{figure}
    \centering
    \begin{tabular}{cc}   
        \begin{minipage}[b]{0.5\linewidth}
            \centering
            \includegraphics[keepaspectratio, scale=0.32]{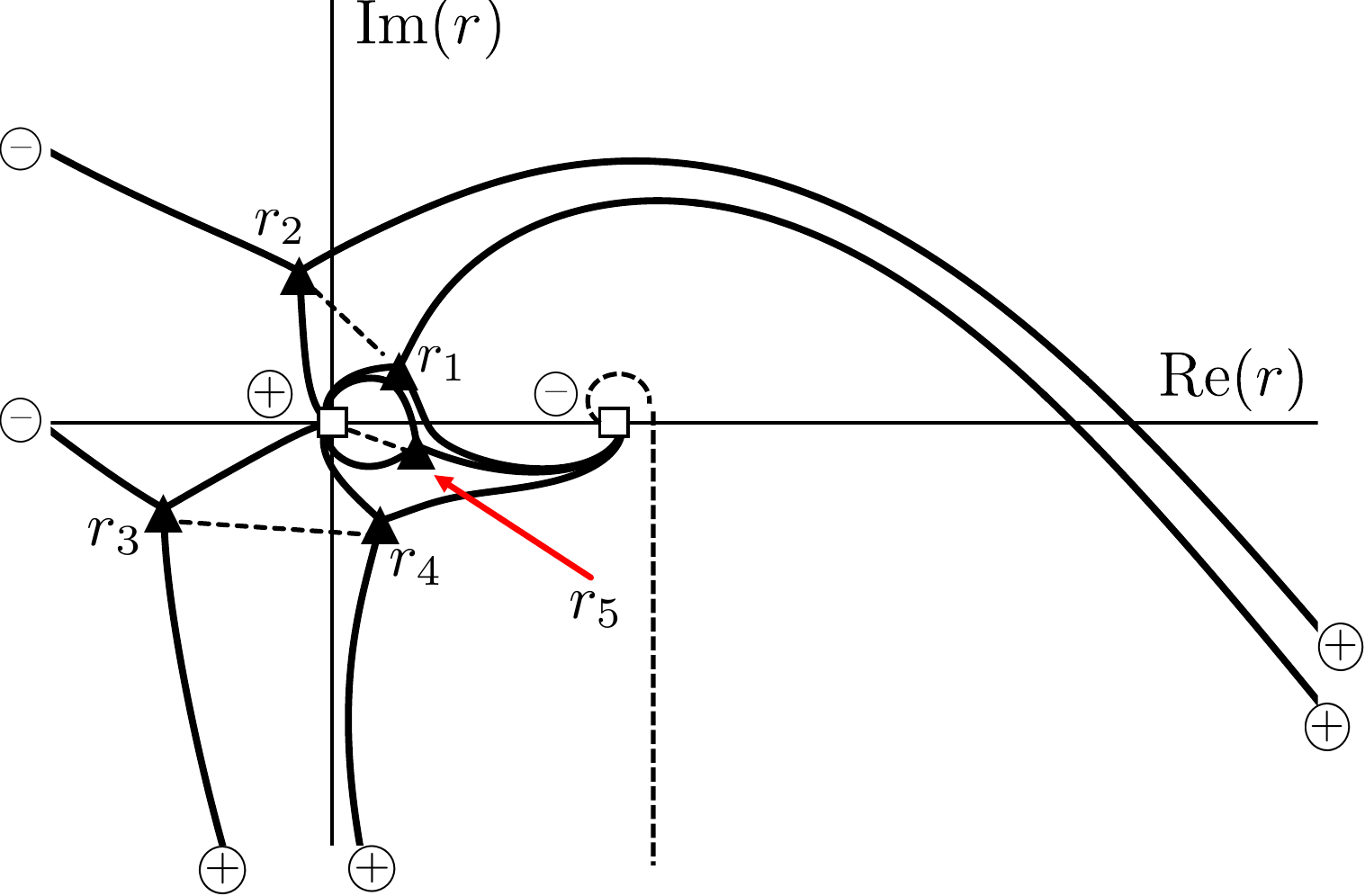}
        \end{minipage}&
        \begin{minipage}[b]{0.5\linewidth}
                \centering
                \includegraphics[keepaspectratio, scale=0.32]{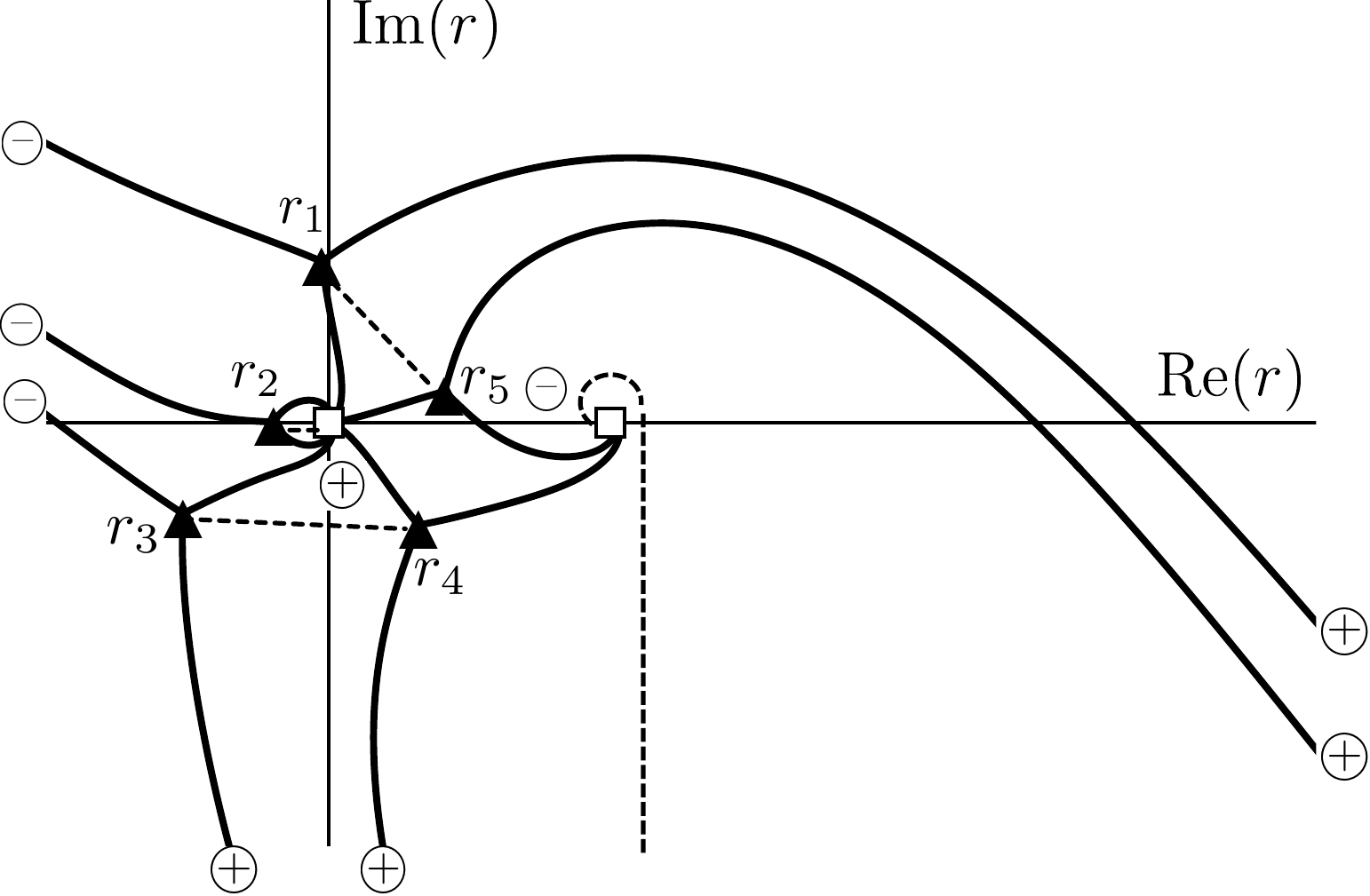}
        \end{minipage}
    \end{tabular}
    \caption{
    Schematic pictures of the Stokes curves for the potential $Q_{0}^{(4)}$ with the branch cuts. We take $\ell=s=2$ and $\omega = 6 - 6 i$ for both panels. The coefficient $\alpha_4$ is set to $\alpha_4=1$ (left) and $\alpha_4=-1$ (right).}
    \label{fig:SL_parametrized_Q4}
\end{figure}

Next, we consider the correction $\delta Q_4$ in the perturbation equation. 
The potential is given by
\begin{align}
    Q^{(4)}_{0}\coloneq
    Q_{\text{RW,0}}+\delta Q_4 
    %= \frac{-\omega^2r^4+(r-1)(Lr-s^2)}{r^2(r-1)^2}
    %+\frac{\alpha_4}{r^3(r-1)}
    =\frac{-\omega^2r^5+(r-1)(Lr^2-s^2r+\alpha_4)}{r^3(r-1)^2}~.
\end{align}
Due to the presence of a higher power of $1/r$, not only the order of the pole at $r=0$ is upraised, but also the numerator is now a quintic polynomial, not quartic as in the Schwartzschild case~\eqref{eq:Q0}. Therefore, the number of turning points increases by one from the Schwartzschild case. 
Then new Stokes curves associated with the new turning point appear, and the structure of the Stokes geometry changes. 
For $\alpha_4 > 0$ (left panel of Fig.~\ref{fig:SL_parametrized_Q4}), a new Stokes curve flowing into the horizon $r = 1$ appears, and for $\alpha_4 < 0$  (right panel of Fig.~\ref{fig:SL_parametrized_Q4}), a new Stokes curve flows to infinity. 

Let us derive the connection matrix $M$ in \eqref{eq:matrix_analytic_continuation} with the $\delta Q_4$ correction. We start with the case of $\alpha_4 > 0$. When performing analytic continuation from $r=1$ to $r=\infty$, where we impose the boundary conditions \eqref{eq:QNMcondition}, along the real axis, we must cross the additional Stokes curve. Taking into account the associated Stokes phenomenon, the connection matrix is now given by
\begin{equation}
    M 
    = \lim_{n\to\infty}\bigg[(V_-)^{-1}\Gamma_{15}(V_-)^{-1}\Gamma_{54}(V_-)^{-1}\Gamma_{41}D_1\bigg]^{n} V_+\Gamma_{12}V_+\Gamma_{21}\,,
\end{equation}
with its components
\begin{align}
\begin{split}
M_{11} &= \lim_{n\to\infty} (\nu_1^+)^n\,,\\
M_{12} &= \lim_{n\to\infty} i \left(1+u_{12}^2\right) (\nu_1^+)^n\,,\\
M_{21} &= \lim_{n\to\infty} \frac{-i \nu_1^+ \left((\nu_1^-)^n-(\nu_1^+)^n\right)}{\nu_1^- -\nu_1^+}
\left( 1 + u_{41}^2 + u_{51}^2 \right)
\,,\\
M_{22} &= \lim_{n\to\infty} \left[(\nu_1^-)^n + \frac{ \nu_1^+ \left((\nu_1^-)^n-(\nu_1^+)^n\right)}{\nu_1^- -\nu_1^+} 
\left( 1 + u_{41}^2 + u_{51}^2 \right)
\left(1+u_{12}^2\right)\right]\,,
\end{split}
\label{m_elements}
\end{align}
Compared to the Schwartzschild case~\eqref{eq:contourmatrix_Schwarzschild}, we find that the additional factor of $(V_{-1})^{-1}\Gamma$ corresponds to the additional Stokes curve from the $\delta Q_4$ correction.
Note that we have used $u_{14}u_{45}=u_{51}^{-1}$ and $u_{45} u_{51} = u_{41}$ in \eqref{m_elements}.
Demanding the appropriate boundary condition of the QNM, $M_{22} = 0$, we obtain the QNM condition as
\begin{align}
    \omega = \frac{1}{4\pi}\log \left( u_{12}^2 +u_{41}^2 + u_{12}^2 u_{41}^2 + u_{51}^2 + u_{52}^2 \right) 
    {-\frac{i}{2} \left( N + \frac{1}{2} \right)}~.
    \label{eq:QNM_condition_deltaQ4_re_p}
\end{align}
In this derivation, we have utilized $\lim_{n \to \infty} \left( \nu_1^+ / \nu_1^- \right)^n \to 0$ due to the same reasoning as described below \eqref{eq:M22_Schwarzschild}.

\begin{figure}
    \centering
    \begin{tabular}{cc}   
        \begin{minipage}[b]{0.5\linewidth}
            \centering
            \includegraphics[keepaspectratio, scale=0.33]{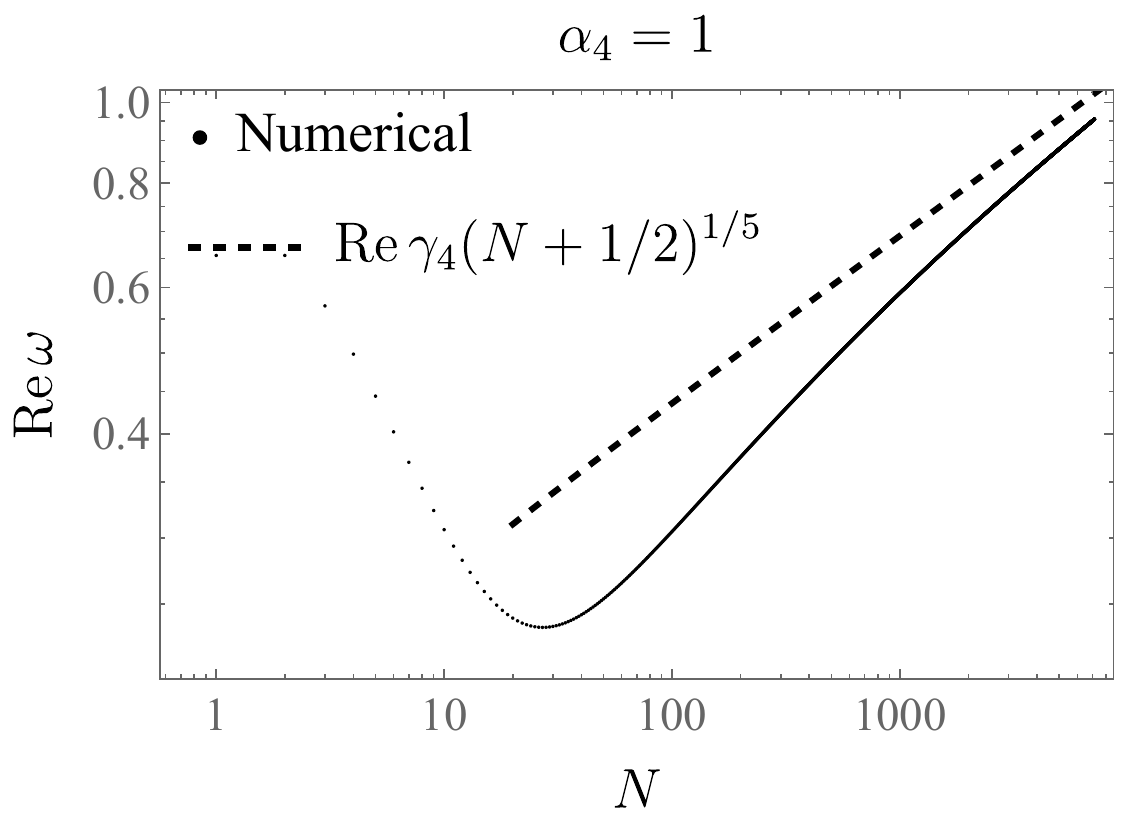}
        \end{minipage}&
        \begin{minipage}[b]{0.5\linewidth}
                \centering
                \includegraphics[keepaspectratio, scale=0.33]{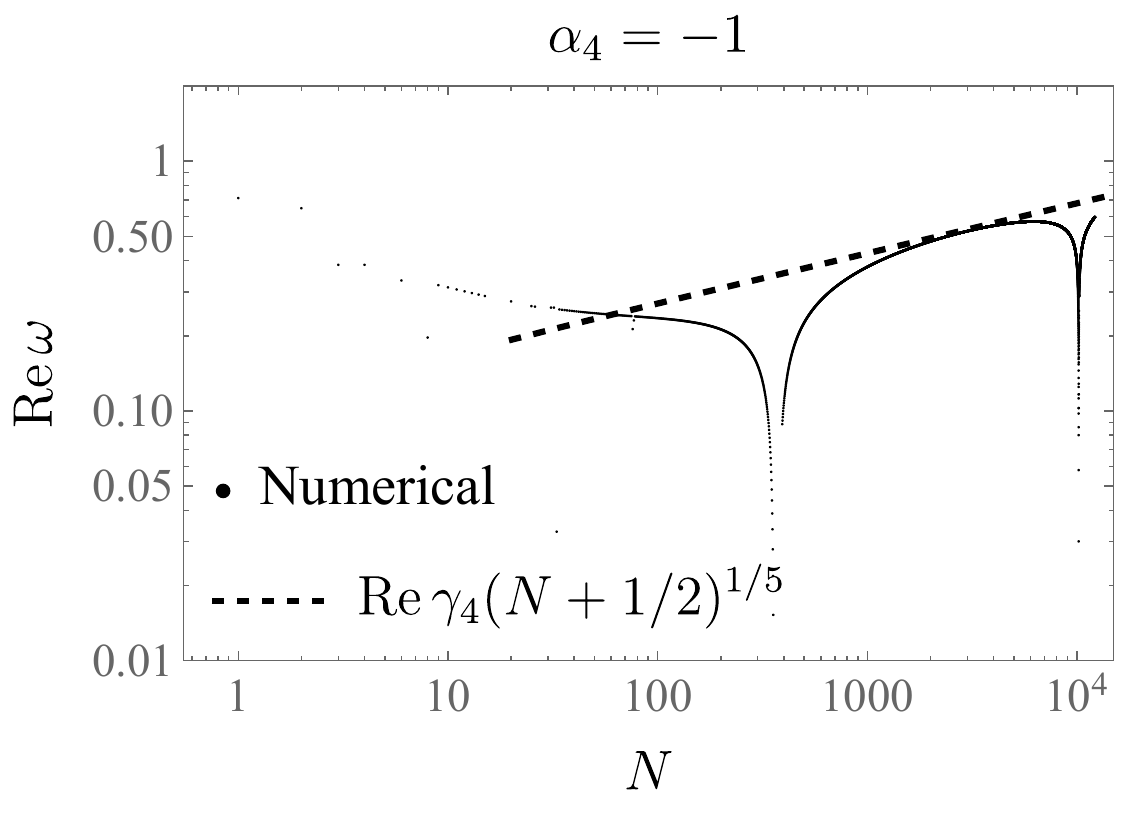}
        \end{minipage}
    \end{tabular}
    \caption{The asymptotic behaviors of $\text{Re}(\omega_N)$ for $\delta Q_4$ with $\alpha_4 = +1$ (left) and $\alpha_4 = -1$ (right) are shown, where we take $\ell = s = 2$. The black dense dots and black dashed line correspond to the numerical result (Leaver) and the prediction of the exact WKB analysis, respectively.}
    \label{fig:Rew_j4_alpha_div}
\end{figure}

For $\alpha_4<0$, no additional Stokes curves cross our analytic continuation path. 
Therefore, the QNM condition is essentially the same as that of the Schwartzschild case in Eq.~\eqref{eq:QNMcondition_re}, but with the replacement of the numbering of the turning points as $r_1\to r_5,\,r_2\to r_1$.
We then arrive at the expression
\begin{align}
    \omega = \frac{1}{4\pi}\log(u_{51}^2+u_{45}^2+u_{51}^2u_{45}^2) 
    {- \frac{i}{2} \left( N + \frac{1}{2} \right)}~.
    \label{eq:QNM_condition_deltaQ4_re_m}
\end{align}
By evaluating the phase integrals $u_{ij}^2 = \exp( 2 I_{ij})$, we obtain, after a straightforward but lengthy computation,
\begin{align}
    I_{ij} \equiv \int_{r_i}^{r_j} S_{\rm odd} \, \dd r
    = i\frac{\sqrt{\pi}\Gamma(-2/5)}{\Gamma(1/10)}\alpha_4^{2/5}(\theta_j^2-\theta_i^2)\omega^{1/5} + \mathcal{O}(\omega^{-1/5})~, \quad \theta_j = \ee^{i\pi(2j-1)/5}~.
\end{align}
Its detailed derivation can be found in Appendix~\ref{app:phase_int_Qp}. 
Note that $\rm{arg}(\omega) \sim -\pi/2$ due to the large and negative $\mathrm{Im}(\omega)$. Thus, the real part of $I_{ij}$ must satisfy
\begin{align}
    \text{Re}\,I_{12} < 0~, \qquad
    \text{Re}\,I_{41} = 0~, \qquad
    \text{Re}\,I_{51} >
    \text{Re}\,I_{52} > 0~,
\end{align}
for $\alpha_4 >0$, and 
\begin{align}
    \text{Re} \, I_{51}>0~, \qquad
    \text{Re} \, I_{45}=0~,
\end{align}
for $\alpha_4 < 0$.
Therefore, the leading behavior in the $\log(\cdots)$ of Eqs.~\eqref{eq:QNM_condition_deltaQ4_re_p} and~\eqref{eq:QNM_condition_deltaQ4_re_m} are both determined by $u_{51}^2$ term. 
As a result, the QNM frequencies in the high-overtone limit are given by
\begin{align}
    \omega  
    &\simeq \gamma_4 {\left( N + \frac{1}{2} \right)^{1/5} }
    {-\frac{i}{2} \left( N + \frac{1}{2} \right)}~, \label{eq:AsymptoticQNM_deltaQ4}
\end{align}
where
\begin{align}
    \gamma_4 = -\frac{\ee^{-i\pi/10}\sin(2\pi/5)}{2^{1/5}\sqrt{\pi}} \, \frac{\Gamma(-2/5)}{\Gamma(1/10)} \, \alpha_4^{2/5}~,
\end{align}
regardless of the sign of $\alpha_4$.
We observe that the real part of QNMs diverges as $N^{1/5}$. 
In Fig.~\ref{fig:Rew_j4_alpha_div}, we compare the numerically obtained QNMs and the analytic result~\eqref{eq:AsymptoticQNM_deltaQ4} for $\alpha_4 = \pm 1$. We clearly see that the numerical result asymptotes to the prediction of the exact WKB analysis in the high-overtone limit. 
It is worth noting that the result \eqref{eq:AsymptoticQNM_deltaQ4} remains valid for small $\alpha_4$, because the Stokes geometry (Fig.~\ref{fig:SL_parametrized_Q4}) does not qualitatively change if we take $\omega$ to be sufficiently large (see below for a further discussion).

With the expression~\eqref{eq:AsymptoticQNM_deltaQ4}, the Schwarzschild result cannot be recovered by taking the $\alpha_4 \to 0$ limit. This is because we implicitly assume the non-zero value of $\alpha_4$ and large $|\omega|$ (more precisely $\alpha_4 |\omega|^{1/2} \gg 1$), and thus the turning points reside at the vertex of the pentagon around $r=0$, resulting from $r^5 \sim -\alpha_4/\omega^2$. 
In the opposite limit, $\alpha_4 |\omega|^{1/2} \ll 1$, one of the turning points, $r = r_5$, becomes much closer to the origin ($r_5 \sim \alpha_4/s^2$) than the other four ($r_i^4 \sim s^2/\omega^2$) with $i =1$, $2$, $3$, or $4$.
In this case, with an appropriate evaluation of the phase integral, we can recover the QNM frequencies of the Schwartzschild black hole in the limit of $\alpha_4 \to 0$. 
Let us focus on $\alpha_4 \to 0^+$ as an example. 
In this case, $r_1$ to $r_4$ in Fig.~\ref{fig:SL_parametrized_Q4} asymptote to the values in the case of the Schwarzschild black hole, whereas $r_5 \sim \alpha_4/s^2$. 
With these turning points, we can show $I_{5j} \to -\infty$ (see Appendix~\ref{app:I_5}). Thus, the phase integrals $u_{51}$ and $u_{52}$ in Eq.~\eqref{eq:QNM_condition_deltaQ4_re_p} vanishes in the limit $\alpha_4 \to 0^+$ and we recover the Schwartzschild QNM condition~\eqref{eq:QNMcondition_re}.

\subsection{Comments on $\delta Q_{p\geq5}$ correction and general corrections}

In the previous subsections, we have studied $\delta Q_3$ and $\delta Q_4$. Here, we provide a general statement for higher-order parametrized corrections $\delta Q_{p\geq 5}$.
In general, the QNMs condition is given by
\begin{align}\label{eq:qnmgen}
    \omega = \frac{1}{4\pi}\log\left[(\text{sum of products of }u_{ij})\right] 
    {- \frac{i}{2} \left( N + \frac{1}{2} \right)} \ .
\end{align}
For large $\omega$, the exponents of $u_{ij}$ generically scale as $\omega^{(p-3)/(p+1)}$, as far as $\alpha_j$ is constant (see Appendix~\ref{app:phase_int_Qp}). This is a consequence of the turning points forming the $(p+1)$-gon on the complex plane, that is, they are determined by the equation $r^{p+1} \sim \alpha_p/\omega^{2}$.
Then, the above condition~\eqref{eq:qnmgen} becomes
\begin{align}
    \omega
    &\simeq C\omega^{\frac{p-3}{p+1}}
    {-\frac{i}{2}(N+1/2)}\notag\\
    &\simeq C' ({N+1/2})^{\frac{p-3}{p+1}} 
    {-\frac{i}{2}(N+1/2)},
    \label{eq:AsymptoticQNM_Qp}
\end{align}
where $C$ and $C'$ are complex valued constants. Therefore, the real parts of QNMs diverge as $N^{\frac{p-3}{p+1}}$ for constant $\alpha_j$ corrections. Other cases, e.g., $\alpha_j$ depending on $\omega$ such as the scalar Gauss-Bonnet theory~\cite{Green:1984sg}, are left for future works.

\section{Conclusion and discussion}
\label{sec:Conclusion}
In this paper, we have analytically investigated the asymptotic quasinormal mode (QNM) spectrum of parametrized black holes, $\omega = \omega_N$, using the exact WKB analysis. 
By systematically incorporating small corrections in the Regge-Wheeler-type equation based on the parametrized QNM formalism, we have demonstrated that these parametrized modifications generically alter the large-overtone structure of QNMs. 
A key finding is that the beyond-GR corrections in the perturbation equation can alter the behavior of high-overtone QNMs as $\displaystyle \lim_{N \to \infty} \text{Re} (\omega_N) = \infty$ while the Schwarzschild spectrum in General Relativity (GR) exhibits a convergence of the real parts of QNMs $\displaystyle \lim_{N \to \infty}\text{Re} (\omega_N) =$ const. in the high-overtone limit.
This observation may affect the possible relation between the quantum-gravitational picture and QNM frequencies \cite{Hod:1998vk,Motl:2002hd,Andersson:2003fh,Keshet:2007nv,Maggiore:2007nq}.
The following discussion is based on Bekenstein's conjecture \cite{Bekenstein:1974jk}, which states that the horizon area of a black hole is quantized in units of the squared Planck length $l_P^2$ (in units where $c=G=1$, $l_P=\hbar^{1/2}$).
For a Schwarzschild black hole, Hod interpreted the real part of the QNM frequency in the large-overtone limit, $\mathrm{Re}(\omega)\to \log 3/(4\pi)$, as the minimal quantum of emitted energy. This leads to the area spacing $\Delta A = 4\log (3)\,l_P^2$ \cite{Hod:1998vk}.
In contrast, Maggiore proposed interpreting QNMs as damped harmonic oscillators and argued that the relevant physical frequency is the proper frequency $\omega_0=\sqrt{(\mathrm{Re}\,\omega)^2+(\mathrm{Im}\,\omega)^2}$ rather than $\mathrm{Re}(\omega)$ \cite{Maggiore:2007nq}.
With this prescription and high-overtone limit $\omega_0\simeq\mathrm{Im}(\omega)$, the area spacing becomes $\Delta A =8\pi l_P^2$ which reproduces Bekenstein's original result \cite{Bekenstein:1974jk}.
Our results in Eqs.~\eqref{eq:qnmlogNdiv} and \eqref{eq:AsymptoticQNM_deltaQ4} suggest that the equally spaced area spectrum is not supported from the viewpoint of Hod's interpretation, whereas it remains consistent within Maggiore's framework.

We have also pointed out (at the end of Sec.~\ref{subsec:deltaQ4}) that taking the Schwarzschild limit from the perturbation potential with the $\delta Q_4$ correction needs an extra care. Simply setting a vanishing value of the parametrization constant, $\alpha_4$ in this case, appearing in the final result leads to an inconsistent Schwarzschild QNM spectrum, at least in the high-overtone regime. This is due to the fact that the expansion parameter for ``high overtones'' is not $\vert \omega \vert$ but rather $\alpha_4^2 \vert \omega \vert$, and thus taking $\alpha_4 \to 0$ is not a consistent treatment. To correctly achieve the Schwarzschild limit, one needs to go back to the analysis of the Stokes geometry for small $\alpha_4$ \textit{before} considering large $\vert \omega \vert$ values. The structure of the Stokes curves in this regime indeed differs from that in the $\alpha_4^2 \vert \omega \vert \gg 0$ one, which, properly taken into account, leads to the correct QNM values of the Schwarzschild. 

Furthermore, our work demonstrates that the exact WKB analysis serves as a useful analytic tool for studying QNMs even in black holes beyond GR. 
Understanding the physical interpretation behind the convergence or divergence of $\text{Re}(\omega_N)$, as well as its possible connection to QNM spectral instability, would be an interesting direction for future investigation.

\acknowledgments
T.~M.~was supported by JSPS KAKENHI Grant Number JP23KJ1543 and in part by the 2025 Osaka Metropolitan University (OMU) Strategic Research Promotion Project (Development of International Research Hubs).
R.~N.~was in part supported by MEXT KAKENHI Grant Number JP23K25868 and by the RIKEN Incentive Research Project grant.
H.~O.~was supported by JSPS KAKENHI Grant Numbers JP23H00110, JP25K17388 and Yamada Science Foundation. 
N.~O.~was supported by Japan Society for the Promotion of Science (JSPS) KAKENHI Grant No.~JP23K13111 and by the Hakubi project at Kyoto University.

\appendix

\section{Calculation of phase integrals}
In this appendix, we calculate the phase integrals $u_{ij}$ appeared in section \ref{sec:Parametrized_BH} and \ref{sec:AsymQNMs}.
For simplicity of notation, reset the symbols defined in each subsection at the beginning of that subsection.

\subsection{$\delta Q_3$}
\label{app:phase_int_Q3}

For the Regge-Wheeler potential \eqref{eq:Q0},
$u_{ij}$ in Section \ref{sec:AsymQNMs} can be obtained by replacing $s^2 \to s^2-\alpha_3$.
We first approximate the integrand in $u_{ij}$ as
\begin{align}
    I_{ij}
    := \frac{1}{2} \oint_{\gamma_{ij}} S_{\text{odd}}~\dd r
    \simeq \int_{r_i}^{r_j}\sqrt{Q_{\text{RW},0}} ~\dd r \; ,
\end{align}
where we take $\eta=1$.
We formally expand $\sqrt{Q_{\text{RW},0}}$ for large $\omega$ as
\begin{align}
    \sqrt{Q_{\text{RW},0}}
    &=i\omega \, \frac{r}{r-1}\sqrt{1-J}
    =i\omega \, \frac{r}{r-1}\sum_{n=0}^{\infty}a_nJ^n\;, 
    \label{eq:expandQ0}
\end{align}
where
\begin{align}
 \label{eq:J}
    J(r)&=\frac{1}{\omega^2}\frac{r-1}{r^4} \left[ Lr-s^2 \right] \,, \qquad 
    a_n =-\frac{\Gamma(n-1/2)}{2\sqrt{\pi}n!}\,.
\end{align}
On the other hand, in the case of higher overtones of the Schwarzschild QNMs, the turning points can be expanded for large $\omega$ as
\begin{align}
    &r_j=b_1 \theta_j \omega^{-1/2}
    +b_2 \theta_j^{2}\omega^{-1}
    +\mathcal{O} \big( \omega^{-3/2} \big) \ , \qquad
   \theta_j=\ee^{i\pi(j-1)/2}, \quad (j=1,2,3,4) \:,
\end{align}
where $b_1 = s^{1/2}$, $b_2 = - \left( L+s^2 \right)/(4s)$, and we take $\arg \omega \simeq -\pi/2$ for $\vert {\rm Im} \, (\omega) \vert \gg \vert {\rm Re} \, (\omega) \vert$ to determine the phase $\theta_j$.

Let us consider integrating \eqref{eq:expandQ0} term by term, substituting $r$ by the turning points $r_j$ into the integrated formula, and expanding it in terms of $1/\omega$.
The integration of the first term in eq.~\eqref{eq:expandQ0}, $A_0$, then reads
\begin{align}
    A_0 & := \int^{r_j} i\omega \, \frac{r}{r-1}\,\dd r 
    = -\pi\omega 
    +ic_1^{(0)}\theta_j^{2}
    +ic_2^{(0)}\theta_j^3\omega^{-1/2}
    +\mathcal{O}\big( \omega^{-1} \big) \ , 
    \label{eq:Coefficients_c00}
\end{align}
where $c_1^{(0)} = -s/2$ and $c_2^{(0)} = \left( 3L-s^2 \right) / \left( 12s^{1/2} \right)$.
For $n\geq 1$, it is sufficient to integrate the following function to obtain the $\mathcal{O}\big(\omega^{-1/2} \big)$ contributions:
\begin{align}
    i \omega \, \frac{r}{r-1} \, a_n J^n
    &= i a_n \, \omega^{1-2n}r^{1-4n} \bigg[d_1^{(n)}+d_2^{(n)}r
    +\mathcal{O}(r^2)\bigg]\;, 
\end{align}
where $d_1^{(n)} = -s^{2n}$ and $d_2^{(n)} = s^{2(n-1)} \left[ nL+(n-1)s^2 \right]$.
The integration of this quantity gives
\begin{align}
    A_n \coloneqq \int^{r_j} i \omega \, \frac{r}{r-1} \, a_n J^n \, \dd r \notag
    =
    ia_n\bigg(c_1^{(n)}\theta_j^{2}+c_2^{(n)}\theta_j^{3}\omega^{-1/2}
    \bigg)
    + \mathcal{O} \big(\omega^{-1} \big)\,,
\end{align}
where $c_1^{(n)} = s/(4n-2)$ and $c_2^{(n)} = - \left( 3L-s^2 \right) / \left[ 4(4n-3)s^{1/2} \right]$.
One now observes that $A_0$ in \eqref{eq:Coefficients_c00} can be included in the above expression of $A_n$.

Since direct computations reveal
\begin{align}
    \sum_{n=0}^{\infty} \frac{a_n}{4n-2} = -\frac{\pi}{4} \ , \qquad
    \sum_{n=0}^{\infty} \frac{a_n}{4n-3} = \frac{\sqrt{\pi} \, \Gamma(-3/4)}{8\Gamma(3/4)}
    = -\frac{\Gamma(1/4)^2}{6\sqrt{2\pi}} \, , 
\end{align}
the summation of $A_n$ over $n\,(=0,1,2,\ldots)$ is given by 
\begin{align}
    I_j
    :=\int^{r_j}\sqrt{Q_{\text{RW},0}}\, \dd r 
    =\sum_{n=0}^{\infty} A_n  
    =-\pi\omega +ic_1\theta_j^{2}
    + i c_2 \theta_j^{3} \omega^{-1/2} 
    +\mathcal{O} \big( \omega^{-1} \big)\,,
    \label{eq:ci_Q3}
\end{align}
where 
\begin{align}
    c_1 = -\frac{\pi s}{4} \ , \qquad 
    c_2 = \frac{\Gamma(1/4)^2}{24\sqrt{2\pi}s^{1/2}}(3L-s^2) \ .
\end{align}
Note that the coefficient of $\omega^{-1}$ vanishes because $\theta_j^4=1$.
Then, $I_{ij}$ is given by
\begin{align}
    I_{ij}
    =I_j-I_i 
    = ic_1(\theta_j^2-\theta_i^2) + ic_2(\theta_j^3-\theta_i^3)\omega^{-1/2} 
    + \mathcal{O}(\omega^{-3/2})
    \:.
    \label{eq:I_ij_deltaQ3}
\end{align}

In \eqref{eq:QNMcondition_re}, we need to calculate $u_{12}^2+u_{41}^2+u_{12}^2u_{41}^2$.
Using \eqref{eq:I_ij_deltaQ3} and taking the Taylor expansion for large $\omega$, we obtain
\begin{align}
    u_{12}^2 &= \ee^{2I_{12}} = \ee^{i\pi \sqrt{s^2-\alpha_3}} \left[ 1+2c_2(1-i)\omega^{-1/2}+\mathcal{O}\big(\omega^{-3/2} \big) \right] \ ,\\
    u_{41}^2 &= \ee^{2I_{41}} = \ee^{-i\pi \sqrt{s^2-\alpha_3}} \left[ 1+2c_2(1+i) \omega^{-1/2} + \mathcal{O} \big(\omega^{-3/2} \big) \right] \ ,\\
    u_{12}^2u_{41}^2 &= 1+ 4c_2\omega^{-1/2}+\mathcal{O}\big(\omega^{-1} \big), \\
    u_{12}^2+u_{41}^2+u_{12}^2u_{41}^2 &= 1+2\cos \pi s + 4 \left[ 1+\sqrt{2}\cos \pi (s-1/4) \right] c_2\omega^{-1/2} +\mathcal{O}\big(\omega^{-1} \big) \ .
    \label{eq:ArgOfLog3}
\end{align}
This concludes the computation of the phase integrals required for the case of $\delta Q_3$ deformation.

\subsection{$\delta Q_{p\geq 4}$}
\label{app:phase_int_Qp}
Next, we calculate the phase integrals for $\delta Q_p, (p\geq4)$ at the leading order in the large $\omega$ expansion.
We formally expand the root of $Q_{0}^{(p)}\coloneqq Q_{\mathrm{RW,0}}+\delta Q_p$ as
\begin{align}
    \sqrt{Q_{0}^{(p)}}
    &=i\omega\frac{r}{r-1}\sum_{n=0}^{\infty}a_n J_p^n,
    \label{eq:expand_deltaQp}
\end{align}
where
\begin{align}
    J_p(r)
    &=\frac{1}{\omega^2}\frac{r-1}{r^{p+1}} \left( Lr^{p-2}-s^2r^{p-3}+\alpha_p \right) \ .
\end{align}
Also, for large $\omega$, the turning points can be written as
\begin{align}
    &r_j=(\alpha_p)^{\frac{1}{p+1}} \theta_j \omega^{-\frac{2}{p+1}}
    +\mathcal{O} \big( \omega^{-\frac{4}{p+1}} \big), \\
   &\theta_j=\ee^{i\pi(2j-1)/(p+1)}, \quad (j=1,2,\ldots,p+1) \:.
\end{align}
Then, we integrate \eqref{eq:expand_deltaQp} term by term, substitute turning points and expand around $1/\omega$.
The first term reads
\begin{align}
    A_0 & := \int^{r_j} i\omega \, \frac{r}{r-1}\,\dd r 
    = -\pi\omega 
    -\frac{i}{2}(\alpha_p)^{\frac{2}{p+1}}\theta_j^{2}\omega^{1-\frac{4}{p+1}}
    +\mathcal{O}\big( \omega^{1-\frac{6}{p+1}} \big) \; .
    \label{eq:Coefficients_c00_deltaQp}
\end{align}
For $n\geq 1$, it is sufficient to integrate the following function to calculate up to $O(\omega^{1-\frac{4}{p+1}})$.
\begin{align}
    i \omega \, \frac{r}{r-1} \, a_n J^n
    &= i a_n \, \omega^{1-2n}r^{1-(p+1)n} \bigg[(-1)^{n-1}\alpha_p^n+\mathcal{O}(r)\bigg]\;,
\end{align}
The integration is given by
\begin{align}
    A_n &\coloneqq \int^{r_j} i \omega \, \frac{r}{r-1} \, a_n J^n \, \dd r 
    =
    ia_n \frac{(\alpha_p)^{\frac{2}{p+1}}}{(p+1)n-2}\theta_j^2\omega^{1-\frac{4}{p+1}}
    + \mathcal{O} \big(\omega^{1-\frac{6}{p+1}} \big)\,.
\end{align}
Thus, \eqref{eq:Coefficients_c00_deltaQp} can be included in this expression with an addition of the first term $-\pi \omega$.

Since we have
\begin{align}
    &\sum_{n=0}^{\infty} \frac{a_n}{(p+1)n-2} = -\frac{\sqrt{\pi}\Gamma\bigg(\frac{p-1}{p+1}\bigg)}{4\Gamma\bigg(\frac{p-1}{p+1}+\frac{1}{2}\bigg)}\;,
\end{align}
the summation of $A_n$ over $n\,(=0,1,2,\ldots)$ is given by
\begin{align}
    I_j
    &=-\pi\omega -i\frac{\sqrt{\pi}\Gamma\bigg(\frac{p-1}{p+1}\bigg)}{4\Gamma\bigg(\frac{p-1}{p+1}+\frac{1}{2}\bigg)}(\alpha_p)^{\frac{2}{p+1}}\theta_j^{2}\omega^{1-\frac{4}{p+1}}
    +\mathcal{O} \big( \omega^{1-\frac{6}{p+1}} \big)\,,
    \label{eq:coefficients_deltaQp}
\end{align}
Then, $I_{ij}$ is given by
\begin{align}
\begin{split}
    I_{ij}
    &=I_j-I_i
    = -i\frac{\sqrt{\pi}\Gamma\bigg(\frac{p-1}{p+1}\bigg)}{4\Gamma\bigg(\frac{p-1}{p+1}+\frac{1}{2}\bigg)}(\alpha_p)^{\frac{2}{p+1}}(\theta_j^2-\theta_i^2)\omega^{1-\frac{4}{p+1}}
    + \mathcal{O}(\omega^{1-\frac{6}{p+1}})\, ,
    \end{split}
    \label{eq:I_ij_deltaQp}
\end{align}
providing the expression for the phase integral in the case of the $\delta Q_{p \ge 4}$ modification.

\section{$I_5$ in the limit $\alpha_4 \to 0^+$}
\label{app:I_5}

In this appendix, we calculate $I_5$ appeared in the correction of $\delta Q_4$ in the Schwarzschild limit, $\alpha_4 \to 0^+$.
In this limit, we have
\begin{align}
    Q_{\text{para,0}} \to Q_{\text{RW,0}}\,, \quad r_5 \to \frac{\alpha_4}{s^2} \ .
\end{align}
Using an approximation in \ref{app:phase_int_Q3}, we obtain
\begin{align}
    I_5 
    &\simeq \int^{r_5}\sqrt{Q_{\text{para},0}}\,dr 
    \simeq \int^{\frac{\alpha_4}{s^2}} \sqrt{Q_{\text{RW},0}}\,dr 
    \simeq \sum_{n=0}^{\infty}\int^{\frac{\alpha_4}{s^2}} ia_n\omega^{1-2n}r^{1-4n}d_1^{(n)}\,dr \notag\\
    &= \frac{s}{2}\bigg[\sqrt{1-\frac{\alpha_4^4\omega^2}{s^{10}}}+\text{arcsinh}\bigg(\frac{s^5}{\alpha_4^2i\omega}\bigg)\bigg].
\end{align}
Therefore, taking $\alpha_4\to0^+$ under $\arg (\omega) \sim -\pi/2$, we obtain $I_5 \to +\infty$, i.e., $I_{5 j} = I_{j} - I_{5} \to - \infty$.

\section{Details of numerical computations}
\label{app:C}

Here, we describe numerical computations of QNMs in this study. We utilize Leaver's method, that is, QNMs are determined from a continued fraction. See~\cite{Leaver:1985ax} for details. 

We consider the master equation~\eqref{eq:masterEq} with the $j$-th order correction $\alpha_j/r^j$ in the potential term, that is,
\begin{align}\label{eq:masterj}
    \left[f^2\frac{\dd^2}{\dd r^2}+f f'\frac{\dd}{\dd r}+\omega^2 - f\bigg[\frac{L}{r^{2}} + \frac{1-s^2}{r^{3}} + \frac{\alpha_j }{r^j}\bigg]\right]\Psi(r) = 0~.
\end{align}
We solve the master equation~\eqref{eq:masterj} with the series expansion of the wave function. Our ansatz is
\begin{align}\label{eq:seriessol}
    \Psi = (r-1)^{- i \omega}r^{2 i \omega}\ee^{i \omega r}\sum_{n=0}^{+\infty} a_n \left(\frac{r - 1}{r}\right)^n~,
\end{align}
which clearly satisfies the ingoing boundary condition at the horizon. At infinity, the ansatz~\eqref{eq:seriessol} satisfies the outgoing boundary condition if the summation uniformly converges for $r \in [1,+\infty]$.
However, the uniform convergence is not guaranteed for an arbitrary $\omega$, but only possible for QNMs.

After substituting the ansatz~\eqref{eq:seriessol} to~\eqref{eq:masterj}, we obtain the $j$-term recurrence relation~\cite{Hatsuda:2023geo}
\begin{align}
    A_n a_{n-1} + B_n a_n + C_n a_{n+1} + \alpha_j \sum_{m=0}^{j-2} D_m a_{n - m} = 0~,
\end{align}
\begin{align}
    A_n &= (n - s - 2i \omega) (n + s - 2 i \omega)~,\\
    B_n &= -2 n^2 + 2 (4 i \omega -1)n  - L + s^2 - 1+ 4 i \omega + 8 \omega^2 ~,\\
    C_n &= (n + 1)(n + 1 - 2 i\omega)~,\\
    D_m &= \frac{(-1)^{m+1} (j-2)!}{m! (j-2-m)!}~.
\end{align}
In the case of $j \ge 4$, we apply the Gaussian elimination technique~\cite{Konoplya:2011qq} to bring $j$-term recurrence relation to the three-term recurrence relation. We formally write the resulting three-term recurrence relation as
\begin{align}\label{eq:3recmod}
    A_n' a_{n-1} + B_n' a_{n} + C_n' a_{n + 1} = 0~,
\end{align}
with $a_{-1} = 0$. The coefficients $A_n',B_n',$ and $C_n'$ are given by the combinations of $A_j,B_j, C_j$ and $D_j$ with $j\le n$. 

The necessary and sufficient condition for the recurrence relation~\eqref{eq:3recmod} to have the solution with the uniform convergence is that the continued fraction
\begin{align}
    -\frac{A'_{1}}{B'_{1}-}\frac{C'_{1}A'_{2}}{B'_{2}-} \cdots~,
\end{align}
converges to $a_1/a_0$~\cite{doi:10.1137/1009002}. In other words, the equality
\begin{align}
    -\frac{B_0'}{C_0'} &= -\frac{A'_{1}}{B'_{1}-}\frac{C'_{1}A'_{2}}{B'_{2}}\cdots~,
\end{align}
must hold. Again, this condition does not hold for any $\omega$, it is only satisfied for the QNM frequencies. 

Furthermore, we can show the following relation holds
\begin{align}\label{eq:downward}
    R_n &= \frac{a_{n+1}}{a_n} = - \frac{A'_{n+1}}{B'_{n+1}-}\frac{C'_{n+1}A'_{n+2}}{B'_{n+2}-}\cdots~,
\end{align}
when $\omega$ is the QNM frequencies. Also, one can obtain 
\begin{align}\label{eq:upward}
    \frac{a_{n+1}}{a_{n}} &= -\frac{C_{n}'}{B_{n}' -}\frac{A_{n}' C_{n-1}'}{B_{n-1}' -} \cdots \frac{A_1' C_0'}{B_0'}~,
\end{align}
 from the three-term recurrence relation. Matching Eq.~\eqref{eq:downward} with Eq.~\eqref{eq:upward}, we arrive at another but equivalent condition for QNMs,
\begin{align}\label{eq:overtonecondition}
     -  \frac{A'_{n+1}}{B'_{n+1}-}\frac{C'_{n+1}A'_{n+2}}{B'_{n+2}-}\cdots = -\frac{B_{n}'}{C_{n}'} +\frac{A_n'}{C_{n}'}\frac{C'_{n-1}}{B'_{n-1}-}\frac{A'_{n}C'_{n-1}}{B'_{n-1}-} \cdots \frac{A_1' C_0'}{B_0'}~.
\end{align}
One can stably find the QNM frequency of the $N$-th overtone with the inverted relation~\eqref{eq:overtonecondition} of $n = N$. 

To numerically construct the continued fraction in~\eqref{eq:overtonecondition}, we iteratively use the recursion relation of $R_n$
\begin{align}
    R_{n-1} &= - \frac{A_n'}{B_n' + C_n' R_n}~,
\end{align}
from large to small $n$. 
We start the recursion with the asymptotic form of $R_n$ given by
\begin{align}
    R_n &= q_0 + \frac{q_1}{\sqrt{n}} + \frac{q_2}{n} + \cdots~,
\end{align}
at some large $n$, in this study we start with $\max(10^3, 10 N)$, when we aim to solve for $N$-th overtone. The coefficients $q_i$ are obtained by solving the three-term recurrence relation order by order in terms of $n$. First few terms for $j = 3$ is \cite{Nollert:1993zz}
\begin{align}
    q_0 &= 1~,\\
    q_1 &= i\sqrt{2 i \omega}~,\\
    q_2 &= -\frac{3}{4} - 2 i \omega~.
\end{align}
In our numerical computation, we expand up to $q_4$.

% Bibliography
\bibliographystyle{JHEP}
\bibliography{manuscript_parametrized}

\end{document}